\documentclass{article}

\usepackage{PRIMEarxiv}

\usepackage[utf8]{inputenc} 
\usepackage[T1]{fontenc}    
\usepackage{hyperref}       
\hypersetup{
    colorlinks=true,
    linkcolor=blue,
    filecolor=cyan,      
    urlcolor=magenta,
    }
\usepackage{url}            
\usepackage{booktabs}       
\usepackage{amsfonts,amssymb,amsmath}       
\usepackage{nicefrac}       
\usepackage{microtype}      
\usepackage{lipsum}
\usepackage{fancyhdr}       
\usepackage{graphicx}       
\usepackage{float}
\usepackage{wrapfig}
\usepackage{xcolor}
\usepackage{soul}
\usepackage[nameinlink]{cleveref}
\graphicspath{{Figures/}}     
\usepackage{subcaption}
\usepackage{caption}
\usepackage{vcell}
\usepackage{colortbl}
\usepackage{makecell}

\usepackage{romannum}
\AtBeginDocument{\pagenumbering{arabic}}

\title{Diffusion Models in Medical Imaging: A Comprehensive Survey}

\author{
	Amirhossein Kazerouni \\
	School of Electrical Engineering \\
	Iran University of Science and Technology \\
	Tehran, Iran\\
	\texttt{amirhossein477@gmail.com} \\ \And
	Ehsan Khodapanah Aghdam \\
	Department of Electrical Engineering \\
	Shahid Beheshti University \\
	Tehran, Iran\\
	\texttt{ehsan.khpaghdam@gmail.com} \\ \And
	Moein Heidari \\
	School of Electrical Engineering \\
	Iran University of Science and Technology \\
	Tehran, Iran\\
    \texttt{moein\_heidari@elec.iust.ac.ir} \\ \And
	Reza Azad \\
	Institute of Imaging and Computer Vision\\ 
	RWTH Aachen University \\
	Aachen, Germany \\ 
    \texttt{azad@lfb.rwth-aachen.de}\And
	Mohsen Fayyaz \\
	Microsoft \\
	Berlin, Germany \\
	\texttt{mohsenfayyaz@microsoft.com} \\ \And
	Ilker Hacihaliloglu \\
	Department of Radiology \\
	Department of Medicine \\
	University of British Columbia \\
	British Columbia, Canada \\ 
    \texttt{ilker.hacihaliloglu@ubc.ca} \\ \And
    Dorit Merhof \\
    Faculty of Informatics and Data Science\\ 
	University of Regensburg \\
	Regensburg, Germany \\ 
    \texttt{dorit.merhof@ur.de} \\
}

\begin{document}
\maketitle

\begin{abstract}
Denoising diffusion models, a class of generative models, have garnered immense interest lately in various deep-learning problems. A diffusion probabilistic model defines a forward diffusion stage where the input data is gradually perturbed over several steps by adding Gaussian noise and then learns to reverse the diffusion process to retrieve the desired noise-free data from noisy data samples. Diffusion models are widely appreciated for their strong mode coverage and quality of the generated samples in spite of their known computational burdens. Capitalizing on the advances in computer vision, the field of medical imaging has also observed a growing interest in diffusion models. With the aim of helping the researcher navigate this profusion, this survey intends to provide a comprehensive overview of diffusion models in the discipline of medical imaging. Specifically, we start with an introduction to the solid theoretical foundation and fundamental concepts behind diffusion models and the three generic diffusion modeling frameworks, namely, diffusion probabilistic models, noise-conditioned score networks, and stochastic differential equations. Then, we provide a systematic taxonomy of diffusion models in the medical domain and propose a multi-perspective categorization based on their application, imaging modality, organ of interest, and algorithms. To this end, we cover extensive applications of diffusion models in the medical domain, including image-to-image translation, reconstruction, registration, classification, segmentation, denoising, 2/3D generation, anomaly detection,  and other medically-related challenges. Furthermore, we emphasize the practical use case of some selected approaches, and then we discuss the limitations of the diffusion models in the medical domain and propose several directions to fulfill the demands of this field. Finally, we gather the overviewed studies with their available open-source implementations at our \href{https://github.com/amirhossein-kz/Awesome-Diffusion-Models-in-Medical-Imaging}{GitHub}\footnote{\url{https://github.com/amirhossein-kz/Awesome-Diffusion-Models-in-Medical-Imaging}}. We aim to update the relevant latest papers within it regularly.
\end{abstract}


\begin{figure*}[t]
	\centering
	\begin{subfigure}[][][c]{0.33\textwidth}
		\includegraphics[width=\textwidth]{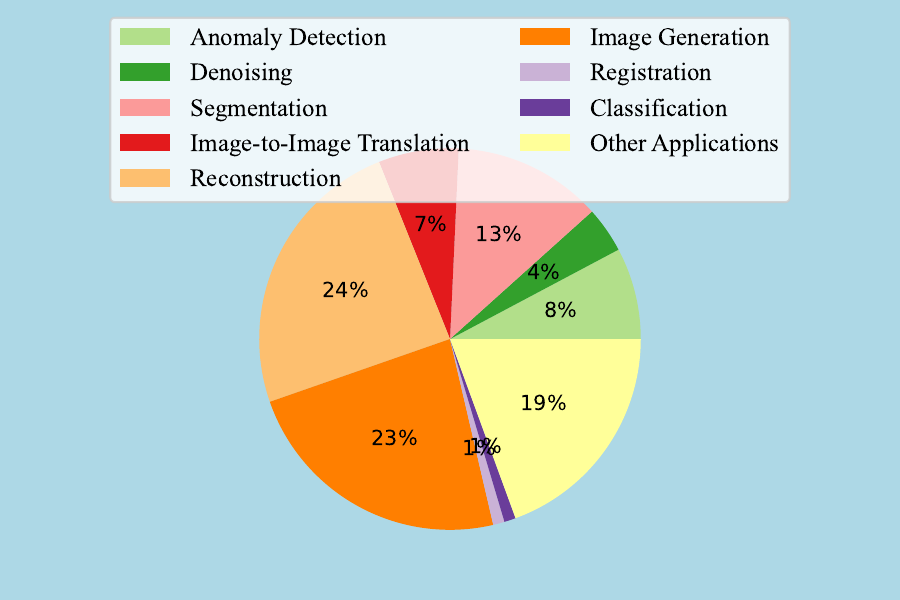}
		\caption{}
		\label{fig:tasks}
	\end{subfigure}
	\hfill
	\begin{subfigure}[][][c]{0.33\textwidth}
		\includegraphics[width=\textwidth]{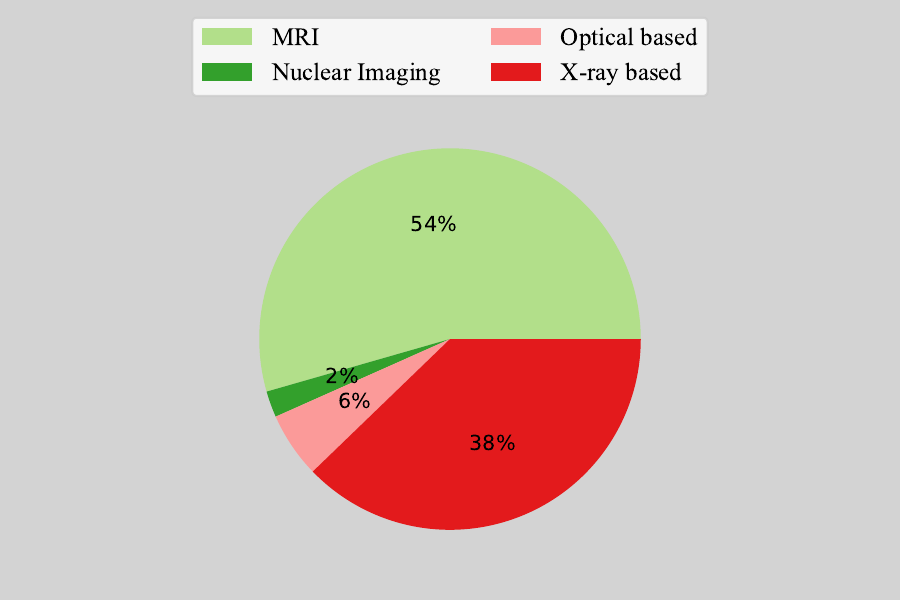}
		\caption{}
		\label{fig:modalities}
	\end{subfigure}
	\hfill
	\begin{subfigure}[][][c]{0.33\textwidth}
		\includegraphics[width=\textwidth]{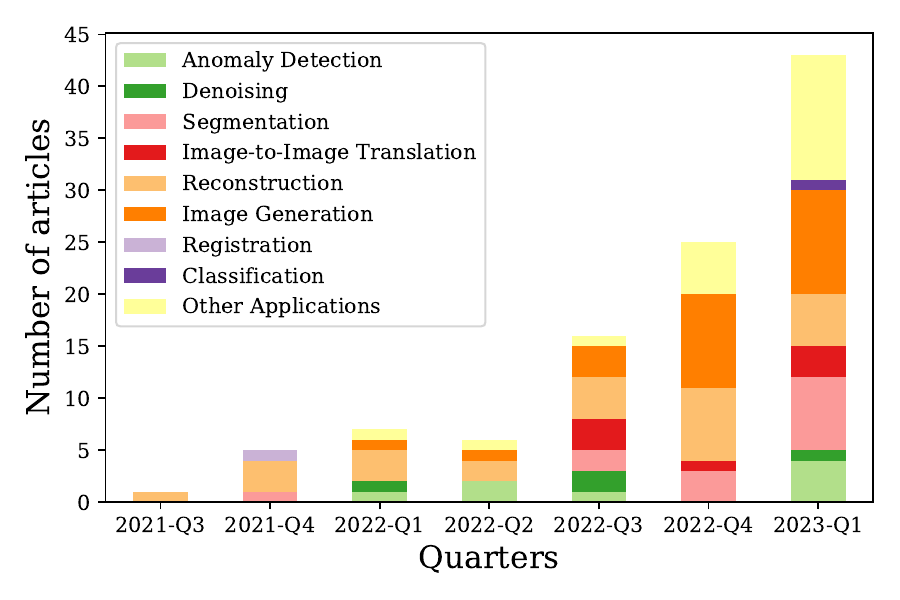}
		\caption{}
		\label{fig:paper-numbers}
	\end{subfigure}
	\caption{The diagram \textbf{(a)} shows the relative proportion of published papers categorized according to their application and \textbf{(b)} according to their imaging modalities. \textbf{(c)} indicates the number of diffusion-based research papers published in the medical domain. The growth rate per year reveals the importance of diffusion models for future work. It is worth mentioning that the overall number of papers is 103.}
\label{fig:charts}
\end{figure*}

\section{Introduction}
\label{intro}
Generative modeling using neural networks has been a leading force in the past decade of deep learning. Since their emergence, generative models have made a tremendous impact in various domains ranging from images \cite{bao2017cvae,razavi2019generating}, audio \cite{kong2020diffwave,oord2016wavenet}, to text \cite{li2022diffusionlm}, and point clouds \cite{yang2019pointflow}. From a probabilistic modeling viewpoint, the key defining characteristic of a generative model is that it is trained in such a way so that its samples $\tilde{\boldsymbol{x}} \sim p_\theta(\tilde{\boldsymbol{x}})$ come from the same distribution as the training data distribution, $\boldsymbol{x} \sim p_d(\boldsymbol{x})$. The energy-based models (EBMs) achieve this by defining an unnormalized probability density over a state space; however, these methods require Markov Chain Monte Carlo (MCMC) sampling during both training and inference, which is a slow iterative process \cite{bond2021deep}. In the past few years, due to the progressions in general deep learning architectures, there has been a resurgence of interest in generative models, unveiling improved visual fidelity and sampling speed. Specifically, generative adversarial networks (GANs) \cite{goodfellow2020generative}, variational autoencoders (VAEs) \cite{rezende2014stochastic}, and normalizing flows \cite{dinh2016density} have emerged. Apart from these, generative models based on diffusion processes offer an alternative to existing VAEs, EBMs, GANs, and normalizing flows, which do not require the alignment of posterior distributions, the estimation of intractable partition functions, the introduction of additional discriminator networks or the placement of network constraints respectively. To date, diffusion models have been found to be useful in a wide variety of areas, ranging from generative modeling tasks such as image generation \cite{dhariwal2021diffusion}, image super-resolution \cite{li2022srdiff}, image inpainting \cite{lugmayr2022repaint} to discriminative tasks such as image segmentation \cite{amit2021segdiff}, classification \cite{zimmermann2021score}, and anomaly detection \cite{wolleb2022diffusionanomaly}. Recently, the medical imaging community has witnessed exponential growth in the number of diffusion-based techniques (see \Cref{fig:charts}). As shown in \Cref{fig:charts}, a wealth of research is dedicated to the applications of diffusion models in diverse medical imaging scenarios. Since diffusion models have recently received significant attention from the research community, the literature is experiencing a large influx of contributions in this direction. Thus, a survey of the existing literature is beneficial for the community and timely. To this end, this survey sets out to provide a comprehensive overview of the recent advances made and provides a holistic overview of this class of models in medical imaging. A thorough search of the relevant literature revealed that we are the first to cover the diffusion-based models exploited in the medical domain. We hope this work will point out new paths, provide a road map for researchers, and inspire further interest in the vision community to leverage the potential of diffusion models in the medical domain. Our major contributions include:

$\bullet$~This is the first survey paper that comprehensively
covers applications of diffusion models in the medical imaging domain. Specifically, we present a
comprehensive overview of all available relevant papers (until October 2022) as well as showcase some of the latest techniques through April 2023.

$\bullet$~We devise a multi-perspective categorization of diffusion models in the medical community, providing a systematical taxonomy of research in diffusion models and their applications. We divide the existing diffusion models into two categories: variational-based models and score-based models. Moreover, we group the applications of diffusion models into nine categories: image-to-image translation, reconstruction, registration, classification, segmentation, denoising, image generation, anomaly detection, and other applications.

$\bullet$~We do not restrict our attention to application and provide a new taxonomy (see \Cref{fig:taxonomy}) where each paper is broadly classified according to the proposed algorithm along with the organ concerned and imaging modality, respectively.

$\bullet$~Finally, we discuss the challenges and open issues and identify the new trends raising open questions about the future development of diffusion models in the medical domain in both algorithms and applications.

\textbf{\emph{Motivation and uniqueness of this survey}.} Generative approaches have undergone significant advances in medical imaging over the past few decades. Therefore, there have been numerous survey papers published on deep generative models for medical imaging \cite{alamir2022role,ali2022role,chen2022generative}. Some of these papers only focus on a specific application, while others concentrate on a specific image modality. There have also been review
articles on diffusion models surfacing recently for computer vision tasks \cite{cao2022survey,yang2022diffusion,croitoru2022diffusion}. Although reviews had already been released before this area is fully developed, a lot of advances in the medical field have come out since then. On the other hand, none of these surveys focuses on the applications of diffusion models in medical imaging, which is the central aspect in pushing this research direction forward. Hence, these surveys leave a clear open gap. Besides, we believe that the medical community can leverage insights from successful products of diffusion models in vision by a retrospective on the past and future research directions of diffusion models provided in our survey. Moreover, diffusion models have demonstrated their potential in generating synthetic data and can serve as an effective supplement to existing real data, as well as a generative prior in biomedical inverse imaging problems (as discussed in \Cref{clinical}). Lastly, we think our survey can help medical researchers (e.g., radiologists) and guide them toward exploiting up-to-date methodologies in their fields. To this end, in this survey, we devise a multi-perspective vision of diffusion models in which we discuss existing literature based on their applications in the medical domain. Nonetheless, we do not restrict our interest to the applications but describe the underlying working principles, the organ, and the imaging modality of the proposed method. We further discuss how this additional information can help researchers attempt to consolidate the literature across the spectrum. A brief outlook of our paper is depicted in \Cref{fig:taxonomy}.

\textbf{\emph{Search Strategy}}. We searched DBLP, Google Scholar, and Arxiv Sanity Preserver with customized search queries, as they allow for customized search queries and provide lists for all scholarly publications: peer-reviewed journal
papers or papers published in the proceedings of conferences or workshops, non-peer-reviewed papers, and preprints. Our search query was \texttt{(diffusion* deep $|$ medical $|$ imaging*) (denoising $|$ medical*) (diffusion* $|$ medical* $|$ probabilistic* $|$ model*) (score* $|$ diffusion* $|$ model* $|$ medical*)}. We filtered our search results to remove false positives and included only papers related to diffusion probabilistic models (e.g., we had many false-positive search results about diffusion Magnetic Resonance Imaging (MRI) models). Notably, we selected the papers for detailed examination based on a careful evaluation of their novelty, contribution, significance, and if being the first introduced paper in medical imaging. After applying these criteria, we selected two or three of the highest-ranked papers to examine in more detail. We acknowledge that there may be other important papers in the field that we did not discuss in our review, but we aimed to provide a comprehensive overview of the most important and impactful papers.

\textbf{\emph{Paper Organization.}} In \Cref{theory}, we present a detailed overview of the concepts and theoretical foundations behind diffusion models, covering two perspectives of diffusion models. In \Cref{clinical}, we will delve into the significance of employing generative models, specifically diffusion models, in clinical settings and discuss the benefits they offer. \Cref{diffusion-action} comprehensively covers the applications of diffusion models in several medical imaging tasks, as shown in \Cref{fig:taxonomy}, and finally provides a comparative task-specific overview of different literature work. We conclude this survey by pinpointing future directions and open challenges facing diffusion models in the medical imaging domain in \Cref{future-direction}.

\begin{figure*}[t]
	\centering
	\begin{subfigure}[][][c]{0.46\textwidth}
		\includegraphics[width=\textwidth]{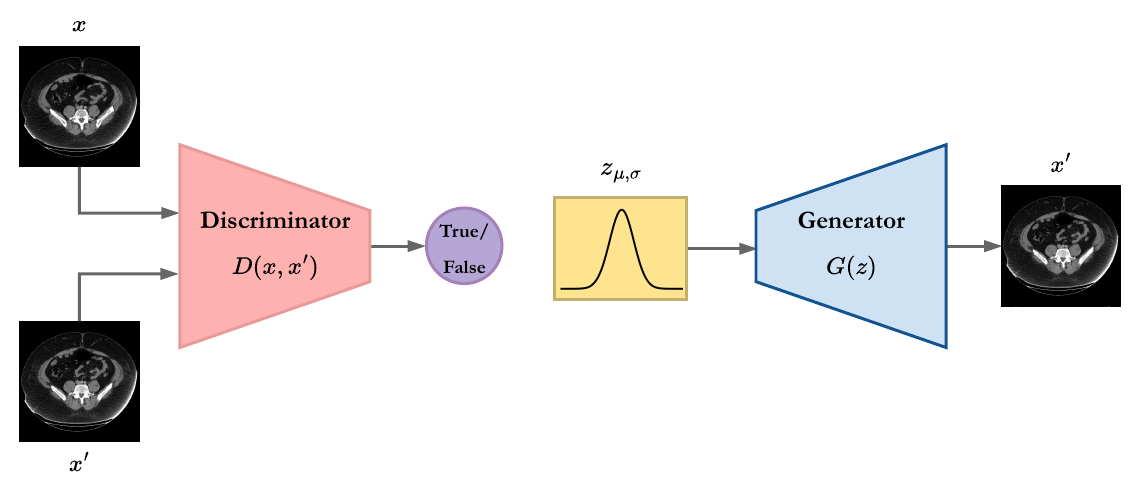}
		\caption{GAN}
		\label{fig:gan}
	\end{subfigure}
	\hfill
	\begin{subfigure}[][][c]{0.46\textwidth}
		\includegraphics[width=\textwidth]{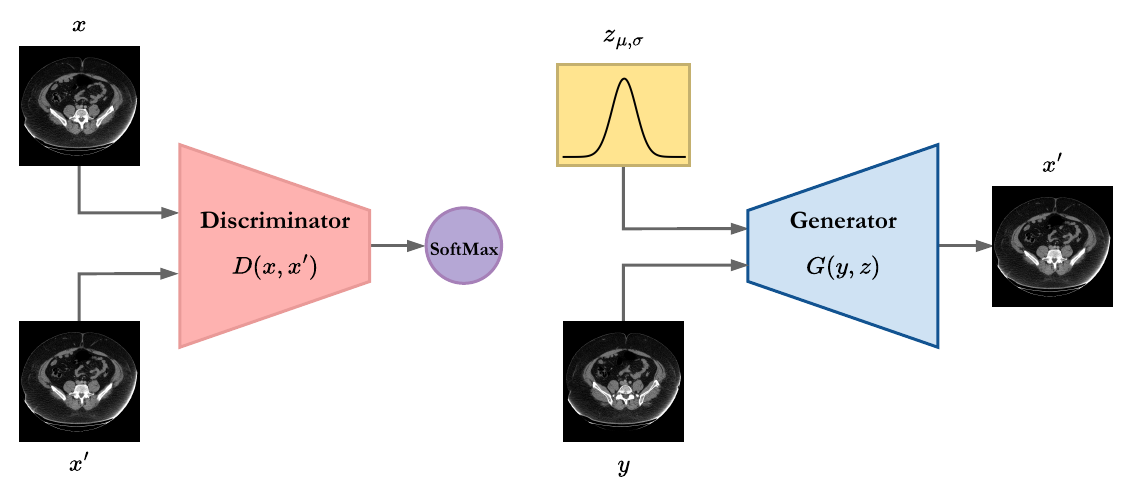}
		\caption{Energy-based Models}
		\label{fig:ebm}
	\end{subfigure}
	\hfill
	\begin{subfigure}[][][c]{0.46\textwidth}
		\includegraphics[width=\textwidth]{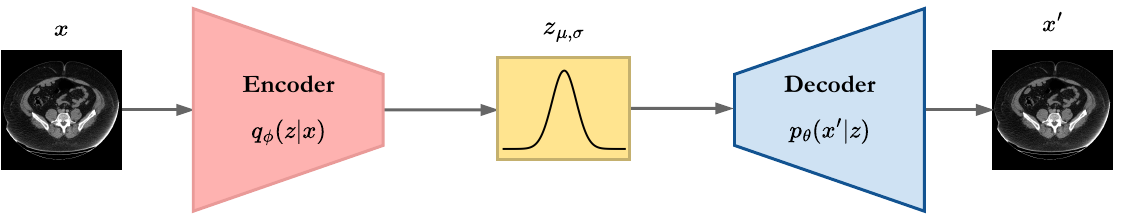}
		\caption{VAE}
		\label{fig:vae}
	\end{subfigure}
	\hfill
	\begin{subfigure}[][][c]{0.46\textwidth}
		\includegraphics[width=\textwidth]{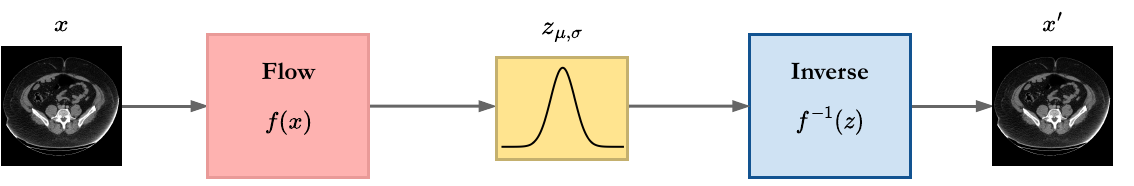}
		\caption{Flow-based models}
		\label{fig:flow}
	\end{subfigure}
	\hfill
	\begin{subfigure}[][][c]{\textwidth}
		\includegraphics[width=\textwidth]{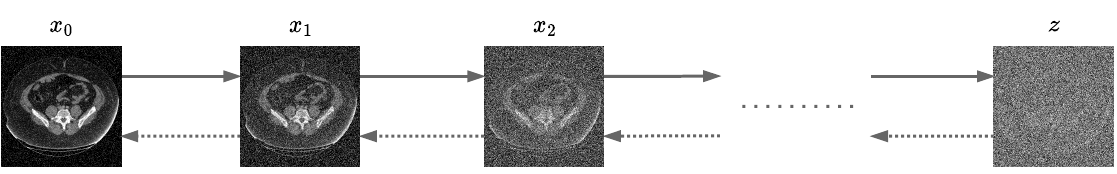}
		\caption{Diffusion Models}
		\label{fig:diffusion}
	\end{subfigure}
	\caption{This figure showcases different generative models and provides an overview of their underlying principles. (\subref{fig:gan}) General Adversarial Network (GAN) \cite{goodfellow2020generative} is an end-to-end pipeline that trains the generator in an adversarial manner to generate samples that the discriminator is capable of distinguishing from the real data sample. (\subref{fig:ebm}) Energy-based Model (EBM) \cite{lecun2006tutorial}, also known as non-normalized probabilistic models, trains in the same way as GANs with two major modifications. First, the discriminator learns a proper energy-based function that maps the data sample to a distribution space. Second, the generator utilizes a prior input to enhance the sample generation performance. (\subref{fig:vae}) Variational AutoEncoder (VAE) \cite{kingma2013auto} is a standalone network that follows a projection from a data sample to a low-dimensional latent space by the encoder and generates by sampling from it via a decoder path. (\subref{fig:flow}) Normalizing flow (NF) \cite{papamakarios2021normalizing} utilizes an invertible flow function to transform input to latent space and generate samples with the inverse flow function. (\subref{fig:diffusion}) Diffusion Models intermingle the noise with the input in successive steps until it becomes a noise distribution before applying a reverse process to neutralize the noise addition in each step in the sampling procedure.}
	\label{fig:generative-pipelines}
\end{figure*}

\section{Theory} \label{theory}
Diffusion models are a cutting-edge class of generative models that have been demonstrated to be highly effective in learning complex data distributions. They are a relatively new addition to the generative learning landscape but have shown to be useful in various applications. In this section, we take an in-depth look at the theory of diffusion models. We begin by discussing the position of diffusion models within the broader generative learning landscape and provide a new perspective on how they compare to other generative models. We further classify diffusion models into two main perspectives: the \textbf{Variational Perspective} and the \textbf{Score Perspective}. We delve into their details and highlight the specific models that fall under them, such as DDPMs in the Variational Perspective and NCSNs and SDEs in the Score Perspective. Ultimately, we provide a comprehensive understanding of the underlying theory behind these methods.

\subsection{Where do diffusion models fit the generative learning landscape?}

\begin{wrapfigure}{R}{0.42\textwidth}
	\centering
	\includegraphics[width=0.4\textwidth]{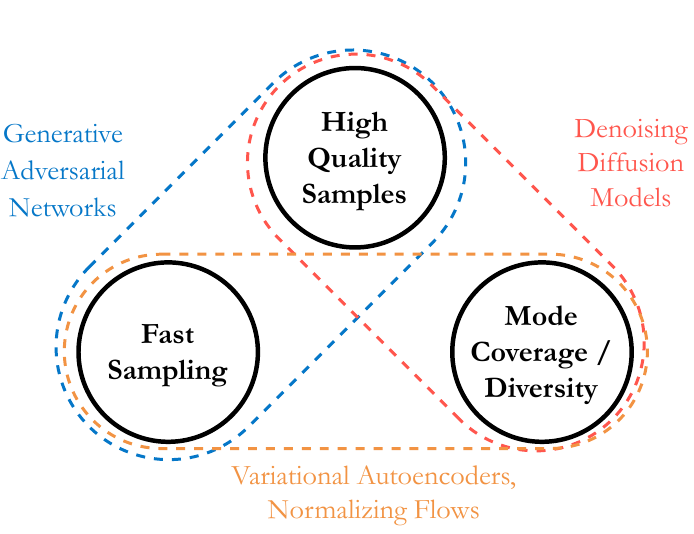}
	\caption{Generative learning trilemma \cite{xiao2022tackling}. Despite the ability of GANs to quickly generate high-fidelity samples, their mode coverage is limited. In addition, VAEs and normalizing flows have been revealed to have a great deal of diversity; however, they generally have poor sampling quality. Diffusion models have emerged to compensate for the deficiency of VAEs and GANs by showing adequate mode coverage and high-quality sampling. Nevertheless, due to their iterative nature, which causes a slow sampling process, they are practically expensive and require more improvement.}
\label{fig:generative-learning}
\end{wrapfigure}

Following the remarkable surge of available datasets, as well as advances in  general deep learning architectures, there has been a revolutionary paradigm shift in generative modeling. Specifically, the three mainstream generative frameworks include, namely, GANs \cite{goodfellow2020generative}, VAEs \cite{rezende2014stochastic,kingma2013auto}, and normalizing flows \cite{dinh2016density} (see \Cref{fig:generative-pipelines}). Generative models typically entail key requirements to be adopted in real-world problems. These requirements include (i) high-quality sampling, (ii) mode coverage and sample diversity, and (iii) fast execution time and computationally inexpensive sampling \cite{xiao2022tackling} (see \Cref{fig:generative-learning}).

Generative models often make accommodations between these criteria. Specifically, GANs are capable of generating high-quality samples rapidly, but they have poor mode coverage and are prone to lack sampling diversity. Conversely, VAEs and normalizing flows suffer from the intrinsic property of low sample quality despite being witnessed in covering data modes. GANs consist of two models: a generator and a critic (discriminator), which compete with each other while simultaneously making each other stronger. The generator tries to capture the distribution of true examples, while the discriminator, which is typically a binary classifier, estimates the probability of a given sample coming from the real dataset. It works as a critic and is optimized to recognize the synthetic samples from the real ones. A common concern with GANs is their training dynamics which have been recognized as being unstable, resulting in deficiencies such as mode collapse, vanishing gradients, and convergence \cite{wiatrak2019stabilizing}. Therefore, an immense interest has also influenced the research direction of GANs to propose more efficient variants \cite{miyato2018spectral,motwani2020novel}. VAEs optimize the log-likelihood of the data by maximizing the evidence lower bound (ELBO). Despite the remarkable achievements, the behavior of VAEs is still far from satisfactory due to some theoretical and practical challenges such as balancing issue \cite{davidson2018hyperspherical} and variable collapse phenomenon \cite{asperti2019variational}. A flow-based generative model is constructed by a sequence of invertible transformations. Specifically, a normalizing flow transforms a simple distribution into a complex one by applying a sequence of invertible transformation functions where one can obtain the desired probability distribution for the final target variable using a change of variables theorem. Unlike GANs and VAEs, these models explicitly learn the data distribution; therefore, their loss function is simply the negative log-likelihood \cite{weng2018flow}. Despite being feasibly designed, these generative models have their specific drawbacks. Since the Likelihood-based method has to construct a normalized probability model, a specific type of architecture must be used (Autoregressive Model, Flow Model), or in the case of VAE, an alternative Loss such as ELBO is not calculated directly for the generated probability distribution. In contrast, the learning process of GANs is inherently unstable due to the nature of the adversarial loss of the GAN. Recently, diffusion models \cite{sohl2015deep,ho2020denoising} have emerged as powerful generative models, showcasing one of the leading topics in computer vision so that researchers and practitioners alike may find it challenging to keep pace with the rate of innovation.

Diffusion models are a powerful class of probabilistic generative models that are used to learn complex data distributions. These models accomplish this by utilizing two key stages: the forward diffusion process and the reverse diffusion process. The forward diffusion process adds noise to the input data, gradually increasing the noise level until the data is transformed into pure Gaussian noise. This process systematically perturbs the structure of the data distribution. The reverse diffusion process, also known as denoising, is then applied to recover the original structure of the data from the perturbed data distribution. This process effectively undoes the degradation caused by the forward diffusion process. The result is a highly flexible and tractable generative model that can accurately model complex data distributions from random noise.
\subsection{Variational Perspective}
The Variational Perspective category includes models that use variational inference to approximate the target distribution, generally by minimizing the Kullback-Leibler divergence between the approximate and target distributions. Denoising Diffusion Probabilistic Models (DDPMs) \cite{sohl2015deep,ho2020denoising} are an example of this type of model, as they use a variational inference approach to estimate the parameters of a diffusion process.

\subsubsection{Denoising Diffusion Probabilistic Models (DDPMs)}

\textbf{Forward Process.} DDPM defines the forward diffusion process as a Markov Chain where Gaussian noise is added in successive steps to obtain a set of noisy samples. Consider $q\left(x_0\right)$ as the uncorrupted (original) data distribution. Given a data sample $x_0 \sim q\left(x_0\right)$, a forward noising process $p$ which produces latent $x_1$ through $x_T$ by adding Gaussian noise at time $t$ is defined as follows:

\begin{equation}
	q\left(x_t \mid x_{t-1}\right)=\mathcal{N}\left(x_t ; \sqrt{1-\beta_t} \cdot x_{t-1}, \beta_t \cdot \mathbf{I}\right), \forall t \in\{1, \ldots, T\},
\end{equation}

where $T$ and $\beta_1, \ldots, \beta_T \in[0,1)$ represent the number of diffusion steps and the variance schedule across diffusion steps, respectively. $\mathbf{I}$ is the identity matrix and $\mathcal{N}(x ; \mu, \sigma)$ represents the normal distribution of mean $\mu$ and covariance $\sigma$. Considering $\alpha_t=1-\beta_t$ and $\bar{\alpha}_t=\prod_{s=0}^t \alpha_s$, one can directly sample an arbitrary step of the noised latent conditioned on the input $x_0$ as follows: 

\begin{equation}
	q\left(\mathbf{x}_t \mid \mathbf{x}_0\right)=N\left(\mathbf{x}_t ; \sqrt{\bar{\alpha}_t} \mathbf{x}_0,\left(1-\bar{\alpha}_t\right) \mathbf{I}\right)
\end{equation}

\begin{equation}
	\mathbf{x}_t=\sqrt{\bar{\alpha}_t} \mathbf{x}_0+\sqrt{1-\overline{\alpha_l}} \epsilon .
\end{equation}

\textbf{Reverse Process.} Leveraging the above definitions, we can approximate a reverse process to get a sample from $q\left(x_0\right)$. To this end, we can parameterize this reverse process by starting at $p\left(\mathbf{x}_T\right)=\mathcal{N}\left(\mathbf{x}_T ; \mathbf{0}, \mathbf{I}\right)$ as follows:
\begin{equation}
	p_\theta\left(\mathbf{x}_{0: T}\right)=p\left(\mathbf{x}_T\right) \prod_{t=1}^T p_\theta\left(\mathbf{x}_{t-1} \mid \mathbf{x}_t\right)
\end{equation}
\begin{equation}
	p_\theta\left(\mathbf{x}_{t-1} \mid \mathbf{x}_t\right)=\mathcal{N}\left(\mathbf{x}_{t-1} ; \mu_\theta\left(\mathbf{x}_t, t\right), \Sigma_\theta\left(\mathbf{x}_t, t\right)\right) .
\end{equation}

To train this model such that $p\left(x_0\right)$ learns the true data distribution $q\left(x_0\right)$, we can optimize the following variational bound on negative log-likelihood:

\begin{equation}
	\label{eq}
	\begin{aligned}
		\mathbb{E}\left[-\log p_\theta\left(\mathbf{x}_0\right)\right] & \leq \mathbb{B}_q\left[-\log \frac{p_\theta\left(\mathbf{x}_{0: T}\right)}{q\left(\mathbf{x}_{1: T} \mid \mathbf{x}_0\right)}\right] \\
		&=\mathbb{E}_q\left[-\log p\left(\mathbf{x}_T\right)-\sum_{t \geq 1} \log \frac{p_\theta\left(\mathbf{x}_{t-1} \mid \mathbf{x}_t\right)}{q\left(\mathbf{x}_t \mid \mathbf{x}_{t-1}\right)}\right] \\
		&=-L_{\mathrm{VL} . \mathrm{B}} .
	\end{aligned}
\end{equation}

Ho et al. \cite{ho2020denoising} found it better not to directly parameterize $\mu_\theta\left(x_t, t\right)$ as a neural network, but instead to train a model $\epsilon_\theta\left(x_t, t\right)$ to predict $\epsilon$. Hence, by reparameterizing \Cref{eq}, they proposed a simplified objective as follows:

\begin{equation}
	L_{\text {simple }}=E_{t, x_0, \epsilon}\left[\left\|\epsilon-\epsilon_\theta\left(x_t, t\right)\right\|^2\right] ,
\end{equation}
where the authors draw a connection between the loss in \Cref{eq} to generative score networks in Song et al. \cite{song2019generative}.

\subsection{Score Perspective}
Score Perspective models rely on a maximum likelihood-based estimation approach, using the score function of the log-likelihood of the data to estimate the parameters of the diffusion process. Noise-conditioned Score Networks (NCSNs) \cite{song2019generative} and Stochastic Differential Equations (SDEs) \cite{song2020score} are both subcategories that fall into this category. NCSNs focus on estimating the derivative of the log density function of the perturbed data distribution at different noise levels, while SDEs are a generalization of previous approaches and encompass both DDPMs and NCSNs characteristics. We hereinafter elaborate on the details of each subcategory.

\subsubsection{Noise Conditioned Score Networks (NCSNs)}
The score function of some data distribution $p(x)$ is defined as the gradient of the log density with respect to the input, $\nabla_x \log p(x)$. To estimate this score function, one can train a shared
neural network with score matching. Specifically, the score network $\mathbf{s}_{\boldsymbol{\theta}}$ is a neural network parameterized by $\boldsymbol{\theta}$, which is trained to approximate the score of $p(x)$ ($s_\theta(x) \approx \nabla_x \log p(x)$) by minimizing the following objective:

\begin{equation}
	\label{ncsn}
	\mathbb{E}_{x \sim p(x)}\left\|s_\theta(x)-\nabla_x \log p(x)\right\|_2^2 .
\end{equation}

However, due to the computational burden of calculating $\nabla_x \log p(x)$, score matching is not scalable to deep networks and high dimensional data. To mitigate this problem, the authors of \cite{song2019generative} propose to exploit denoising score matching \cite{vincent2011connection} and sliced score matching \cite{song2020sliced}. Moreover, Song et al. \cite{song2019generative} highlight major challenges that prevent a naive application of score-based generative modeling in real data. The key challenge is the fact that the estimated score functions are inaccurate in low-density regions since data in the real world tend to concentrate on low-dimensional manifolds embedded in a high-dimensional space (the manifold hypothesis). The authors demonstrate that these problems can be addressed by perturbing the data with Gaussian noise at different scales, as it makes the data distribution more amenable to score-based generative modeling. They propose to estimate the score corresponding to all noise levels by training a single noise-conditioned score network (NCSN). They derive $\nabla_x \log \left(p_{\sigma_t}(x)\right)$ as $\nabla_{x_t} \log p_{\sigma_t}\left(x_t \mid x\right)=-\frac{x_t-x}{\sigma_t}$ by choosing the noise distribution to be $p_{\sigma_t}\left(x_t \mid x\right)=\mathcal{N}\left(x_t ; x, \sigma_t^2 \cdot \mathbf{I}\right)$ where $x_t$ is a noised version of $x$. Thus, for a given sequence of Gaussian noise scales $\sigma_1<\sigma_2<\cdots<\sigma_T$, \Cref{ncsn} can be written as:

\begin{equation}
	\frac{1}{T} \sum_{t=1}^T \lambda\left(\sigma_t\right) \mathbb{E}_{p(x)} \mathbb{E}_{x_t \sim p_{\sigma_t}\left(x_t \mid x\right)}\left\|s_\theta\left(x_t, \sigma_t\right)+\frac{x_t-x}{\sigma_t}\right\|_2^2 ,
\end{equation}
where $\lambda\left(\sigma_t\right)$ is a weighting function. The inference is done using an iterative procedure called "Langevin dynamics" \cite{parisi1981correlation,grenander1994representations}. Langevin dynamics design an MCMC procedure to sample from a distribution $p(\mathbf{x})$ using only its score function $\nabla_{\mathbf{x}} \log p(\mathbf{x})$. Specifically, to move from
a random sample $\mathbf{x}_0 \sim \pi(\mathbf{x})$ towards samples from $p(\mathbf{x})$, it iterates the following:

\begin{equation}
	\label{Langevin-eq}
	x_i=x_{i-1}+\frac{\gamma}{2} \nabla_x \log p(x)+\sqrt{\gamma} \cdot \omega_i ,
\end{equation}
where $\omega_i \sim \mathcal{N}(0, \mathbf{I})$, and $i \in\{1, \ldots, N\}$. When $\gamma \rightarrow 0$ and $N \rightarrow \infty,  \mathbf{x}_i$ samples obtained from this procedure converge to a sample from $p(\mathbf{x})$. The authors of \cite{song2019generative} propose a modification of this algorithm nomenclature as the annealed Langevin dynamics algorithm since the noise scale $\sigma_i$ decreases (anneals) gradually over time to mitigate some pitfalls and failure modes of score matching \cite{song2020improved}.

\subsubsection{Stochastic Differential Equations (SDEs)}
Similar to the aforementioned two approaches, score-based generative models (SGMs) \cite{song2020score} transform the data distribution $q\left(x_0\right)$ into noise. However, by generalizing the number of noise scales to infinity, one can view the previous probabilistic models as a discretization of an SGM. We know that many stochastic processes, such as the diffusion process, are the solution to a stochastic differential equation (SDE) in the following form:
\begin{equation}
	\label{sde-base}
	\mathrm{d} \mathbf{x}=\mathbf{f}(\mathbf{x}, t) \mathrm{d} t+g(t) \mathrm{d} \mathbf{w} ,
\end{equation}
where $\mathbf{f}(., t)$ is the drift coefficient of the SDE, $\mathrm{g}(\mathrm{t})$ is the diffusion coefficient, and w represents the standard Brownian motion.
Let $\mathbf{x}_0$ be the uncorrupted data sample, and $\mathbf{x}_T$ denote the perturbed data approximating standard Gaussian distribution. For the given forward SDE, there exists a reverse time SDE running backward where, by starting with a sample from $p_T$ and reversing this diffusion SDE, we can obtain samples from our data distribution $p_0$. The reverse-time SDE is given as:
\begin{equation}
	\label{sdebackward}
	d \mathbf{x}=\left[\mathbf{f}(\mathbf{x}, t)-g^2(t) \textcolor{red}{\nabla_x \log p_t(x)}\right] d t+g(t) d \bar{\mathbf{w}} ,
\end{equation}
where $d t$ is the infinitesimal negative time step, and $\bar{w}$ is the Brownian motion running backward. In order to numerically solve the reverse-time SDE, one can train a neural network to approximate the actual score function via score matching \cite{song2019generative,song2020score} to estimate $\boldsymbol{s}_\theta(\boldsymbol{x}, t) \simeq \nabla_{\boldsymbol{x}} \log p_t(\boldsymbol{x})$ (denoted \textcolor{red}{red} in \Cref{sdebackward}).
This score model is trained using the following objective:

\begin{footnotesize}
\begin{equation}
	\mathcal{L}(\theta)=\mathbb{E}_{\mathbf{x}(t) \sim p(\mathbf{x}(t) \mid \mathbf{x}(0)), \mathbf{x}(0) \sim p_{\text {data }}}\left[\frac{\lambda(t)}{2}\left\|s_\theta(\mathbf{x}(t), t)-\nabla_{\mathbf{x}(t)} \log p_t(\mathbf{x}(t) \mid \mathbf{x}(0))\right\|_2^2\right] ,
\end{equation}
\end{footnotesize}where $\lambda$ is a weighting function, and $t \sim \mathcal{U}([0, T])$. Notably, $\nabla_{\boldsymbol{x}} \log p_t(\boldsymbol{x})$ is replaced with $\nabla_{\boldsymbol{x}} \log p_{0 t}(\boldsymbol{x}(t) \mid \boldsymbol{x}(0))$ to circumvent technical difficulties.

The sampling process of SDEs can be accomplished by applying any numerical method to \Cref{sdebackward}. Three commonly used techniques are discussed in detail below.
\begin{enumerate}
\label{SDE-sampling-methods}
    \item \textbf{Euler-Maruyama (EM) method:} Using a simple discretization technique in which $dt$ is replaced with $\Delta t$ and $d\bar{w}$ with Gaussian noise $z\sim \mathcal{N}(0, \Delta t\cdot I)$, \Cref{sdebackward} can be solved.
    \item \textbf{Prediction-Correction (PC) method:} In this method, the prediction and correction process takes place in a nested loop, in which the prior data is first predicted and then corrected in several steps. The predictor can be solved using EM. Since the corrector can be any score-based Markov Chain Monte Carlo (MCMC) method, including annealed Langevin dynamics, it can be solved utilizing Langevin dynamics in \Cref{Langevin-eq}.
    \item \textbf{Probability Flow ODE (ODE) method:} SDE equations in \Cref{sde-base} can be written as ODE equations as follows: 
    \begin{equation}
    d \mathbf{x}=\left[\mathbf{f}(\mathbf{x}, t)-\frac{1}{2} g^2(t) \nabla_x \log p_t(x)\right] dt .
    \end{equation}
    \\
    Hence, by solving the ODE problem, $x_0$ can be found. However, while ODE is a quick solver, it lacks a stochastic term to correct errors, resulting in slightly diminished performance.
\end{enumerate}

\section{Clinical Importance}
\label{clinical}
\begin{figure*}[!t]
	\begin{center}
		\includegraphics[scale=0.78]{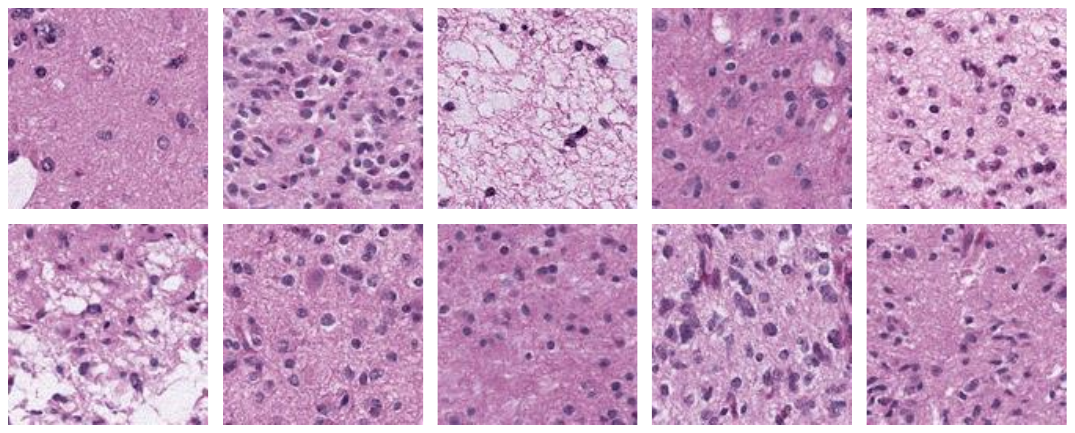}
		\caption{Ten synthetic histopathology images generated by MFDPM \cite{moghadam2023morphology}.}
		\label{fig:MFDPM-output}
	\end{center}
\end{figure*}

Generative models have significantly impacted the field of medical imaging, where there is a strong need for tools to improve the routines of clinicians and patients. Concretely, the complexity of data collection procedures, the lack of experts, privacy concerns, and the compulsory requirement of authorization from patients create a major bottleneck in the annotation process in medical imaging. This is where generative models become advantageous \cite{liu20223d}. Several perspectives have driven our interest in generative diffusion models for medical imaging. In the medical field, many datasets suffer from severe class imbalance due to the rare nature of some pathologies. Diffusion models can alleviate this restriction by generating diverse realistic-looking images to leverage in the medical field. Furthermore, generating synthetic medical images has substantial educational value. With its ability to produce a limitless source of unique instances of different medical imaging modalities, diffusion models can satisfy educational demands by constructing distinct synthetic samples for teaching and practice. Additionally, these artificial images can mitigate data security concerns associated with using patient data in public settings. These artificial images can also solve a particular significant difficulty in training deep neural networks for medical applications. Generally, the annotation of medical images is a lengthy and costly process that necessitates the assistance of an expert. Hence, using diffusion models to generate synthetic samples can alleviate the problem of medical data scarcity to a great extent. A case study using \cite{moghadam2023morphology} to generate histopathology images with rare cancer subtypes is described in \Cref{fig:MFDPM-output}.

While independent use of synthetic data generated by generative models is still in its early stages, studies have shown promising results in utilizing them in real-world scenarios. Studies such as Goncalves et al. \cite{goncalves2020generation} have evaluated different generative methods for creating synthetic electronic health records and found that some have the potential to be useful in practice as they produce synthetic samples that have similar statistical properties to real data without compromising patient privacy. In another study, Chen et al. \cite{chen2021synthetic} found that using both synthetic and real data to train classifiers for histology images improves performance compared to using only real data. Furthermore, studies conducted by Akrout et al. \cite{akrout2023diffusion} have demonstrated that the utilization of synthetic images generated by diffusion models improves the accuracy of skin classifiers, and models that are trained using a combination of synthetic and real data perform better than those trained using only one data source. Moghadam et al. \cite{moghadam2023morphology} conducted a study to assess the morphological properties of synthetic and actual images by administering a survey to two pathologists with different levels of expertise. The results revealed that the pathologists could not distinguish real from synthetic images generated by the diffusion model, and the majority of the small percentage they correctly identified had lower confidence levels. Overall, the research findings indicate that the synthetic images are convincingly similar to real images and can be effective in training models for medical research. In addition, incorporating both synthetic and real data has the potential to improve performance in a variety of applications, as synthetic data serves as a powerful augmentation to real data.

The use of generative models, particularly diffusion models, as a generative prior in biomedical inverse imaging problems is a recent development in the field. In inverse imaging problems, the goal is to infer the underlying physical properties of an object or system from observations or measurements. Conventional methods for inverse imaging problems typically use model-based priors, which are assumptions about the properties of the object or system that are used to guide the reconstruction process. Generative models, specifically diffusion models, can be used as an alternative to these model-based priors since they can provide a more accurate representation of the data distribution \cite{jalal2021robust,chung2022score}. This is because generative models are trained on real data, which means that they can learn the complex patterns and structures present in the data. As a result, they can provide a more accurate prior for the reconstruction process, leading to more accurate reconstructions of the object or system. There are other reasons why diffusion models are proliferating in the field of biomedical inverse imaging problems. One reason is that they can be used to handle high-dimensional and complex data, such as medical images, which are difficult to model using conventional methods. Furthermore, diffusion models can also provide a more efficient and accurate way to infer the underlying physical properties of the object or system and can handle uncertainty and noise in the measurements or observations.

In conclusion, diffusion models have proven to be a valuable and versatile tool that can be utilized in clinical settings and address a wide range of imaging challenges, and it is expected that their use will continue to expand in the future, providing new opportunities for medical imaging and research.

\begin{figure*}
	\centering
	\includegraphics[width=\textwidth]{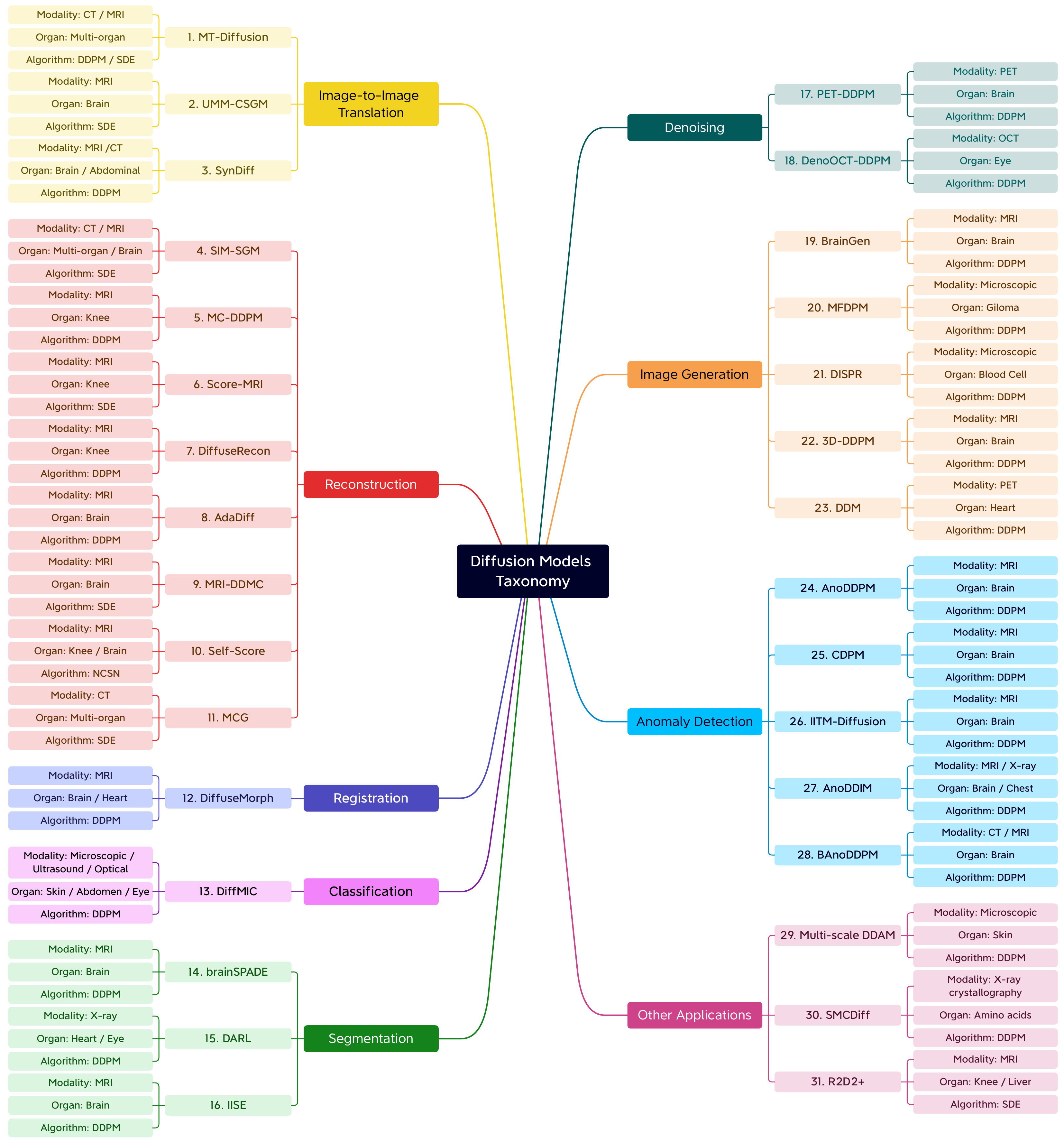}
	\caption{The proposed taxonomy for diffusion-based medical imaging research is built on nine sub-fields: \Romannum{1}) Image-to-Image Translation, \Romannum{2}) Image Reconstruction, \Romannum{3}) Image Registration, \Romannum{4}) Image Classification, \Romannum{5}) Image Segmentation, \Romannum{6}) Image Denoising, \Romannum{7}) Image Generation, \Romannum{8}) Anomaly Detection, and \Romannum{9}) Multi-disciplinary applications, named Other Applications. For the sake of brevity, we utilize the prefix numbers in the paper's name in ascending order and denote the reference for each study as follows:
    1. \cite{lyu2022conversion}, 
    2. \cite{meng2022novel}, 
    3. \cite{ozbey2022unsupervised}, 
    4. \cite{song2022solving}, 
    5. \cite{xie2022measurement}, 
    6. \cite{chung2022score}, 
    7. \cite{peng2022towards}, 
    8. \cite{dar2022adaptive}, 
    9. \cite{luo2022mri}, 
    10. \cite{cui2022self}, 
    11. \cite{chung2022improving}, 
    12. \cite{kim2022diffusemorph}, 
    13. \cite{yang2023diffmic},
    14. \cite{fernandez2022can}, 
    15. \cite{kim2022vessel}, 
    16. \cite{wolleb2022diffusion}, 
    17. \cite{gong2022pet}, 
    18. \cite{hu2022unsupervised}, 
    19. \cite{pinaya2022brain}, 
    20. \cite{moghadam2023morphology}, 
    21. \cite{waibel2022diffusion}, 
    22. \cite{dorjsembe2022three}, 
    23. \cite{kim2022diffusion}, 
    24. \cite{wyatt2022anoddpm}, 
    25. \cite{sanchez2022healthy}, 
    26. \cite{wolleb2022swiss}, 
    27. \cite{wolleb2022diffusionanomaly}, 
    28. \cite{pinaya2022fast},
    29. \cite{wang2022fight}, 
    30. \cite{trippe2023diffusion}, 
    31. \cite{chung2022mr}.} \label{fig:taxonomy}
\end{figure*}

\section{Diffusion Models in Action}
\label{diffusion-action}
Providing a taxonomy for diffusion models more or less follows the same route as other techniques for medical imaging analysis. We provide, however, detailed additional information for each sub-category paper in \Cref{fig:taxonomy}. In this section, we explore diffusion-based methods, which are proposed to solve any disentanglement from the medical imaging analysis in \textbf{seven} application categories, as in 
\Cref{fig:taxonomy}: \textbf{(\Romannum{1})}~Image-to-Image Translation, \textbf{(\Romannum{2})}~Image Reconstruction, \textbf{(\Romannum{3})}~Image Registration, \textbf{(\Romannum{4})}~Image Classification, \textbf{(\Romannum{5})}~Image Segmentation, \textbf{(\Romannum{6})}~Image Denoising, \textbf{(\Romannum{7})}~Image Generation,
 \textbf{(\Romannum{8})}~Anomaly Detection,
and \textbf{(\Romannum{9})}~multi-disciplinary applications, named Other Applications. \Cref{fig:taxonomy} represents a collection of numerous studies for each category with extensive information on each study, such as the modality of study, the organ of interest, and the specific algorithm utilized in the reverse process of the diffusion model for study. Finally, in \Cref{overview}, we discuss the overall algorithms used in the studies and try to shed light on the main novelty and contribution of the papers in \Cref{tab:paperhighlights}.

\subsection{Image-to-Image Translation} \label{sec:translation}
Acquiring multi-modality images for diagnosis and therapy is often crucial. Also, we may miss modalities in some conditions. Diffusion models have shown favorable results for generating missing modalities utilizing cross-modalities and producing ones using other modality types, e.g., translating from MRI to Computed Tomography (CT).

\begin{figure*}[!h]
	\centering
	\includegraphics[width=\textwidth]{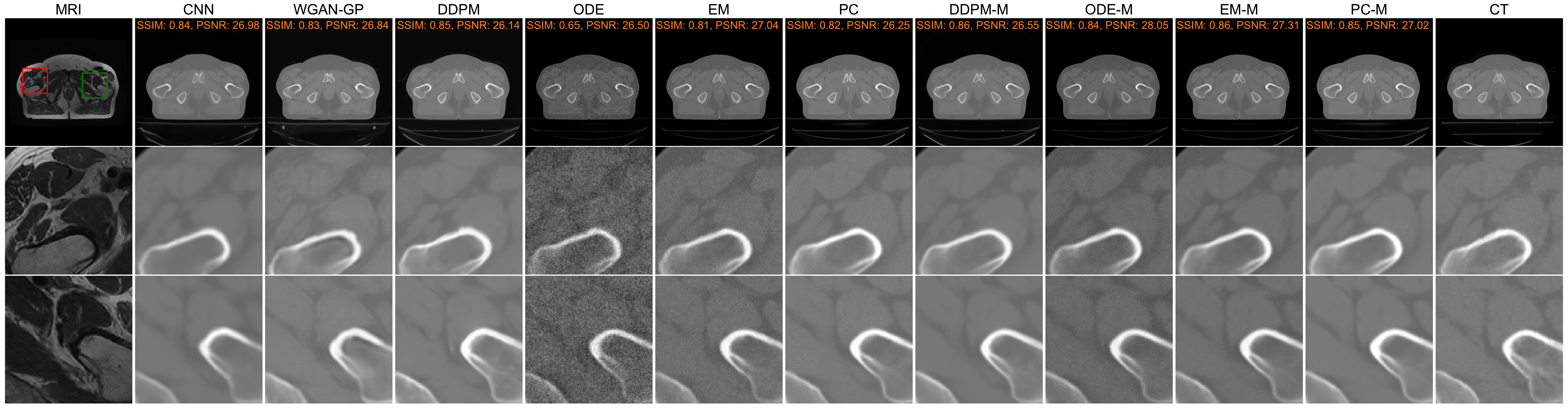}
	\caption{Visual and quantitative comparison of different methods of translating an MR image to a CT image conducted by \cite{lyu2022conversion}. The second and third rows indicate zoomed regions in the original image, delineated by red and green boxes. Results after applying the Monte Carlo method (taking an average from ten outputs) using DDPM, ODE, EM, and ODE sampling methods are displayed with DDPM-M, ODE-M, EM-M, and PC-M, respectively.}
	\label{fig:MRI2CT}
\end{figure*}

CT and MRI are two of the most prevalent imaging types. CT, however, is limited in displaying the intricacies of the images for soft tissue injuries. Hence, a subsequent MRI may be needed for a conclusive diagnosis after receiving the initial CT results. Nevertheless, in addition to being time-consuming and costly, this process may also cause misalignment between MRI and CT images. To this end, Lyu et al. \cite{lyu2022conversion} take advantage of the recently introduced DDPMs \cite{saharia2022image,ho2020denoising} and score-based diffusion models \cite{song2020score} in solving the translation problem between two modalities, i.e., from MRI to CT. In particular, they present conditional DDPM and conditional SDE, in which their reverse process is conditioned on T2w MRI images and conducts comprehensive experiments.
The authors adopt the DDPM and SDE with three different sampling methods (EM, PC, and ODE) that are explained in \Cref{SDE-sampling-methods} and compare their results with the existing GAN-based \cite{gulrajani2017improved} and CNN-based \cite{ronneberger2015u} methods. Their extensive experiments on the Gold Atlas male pelvis dataset \cite{nyholm2018mr} demonstrate that diffusion models outperform both CNN and GAN-based methods in terms of Structural Similarity Index Measure (SSIM) and Peak Signal-to-Noise Ratio (PSNR). Additionally, they employ the Monte Carlo (MC) method to investigate the uncertainties of diffusion models; in this technique, the model outputs ten times, and the average yields the final result. Qualitative results are depicted in \Cref{fig:MRI2CT}.

To cope with the missing modality issue, Meng et al. \cite{meng2022novel} propose a unified multi-modal conditional score-based generative approach (UMM-CSGM), which synthesizes the missing modality based on all remaining modalities as conditions. The proposed model is a conditional format of SDE \cite{song2020score}, in which it employs only a score-based network to learn different cross-modal conditional distributions. Experiments on the BraTS19 dataset \cite{bakas2017advancing,bakas2018identifying,menze2014multimodal}, which contains four MRI modalities for each subject, show that the UMM-CSGM is capable of generating missing-modality images with higher fidelity and structural information of the brain tissue compared to SOTA methods \cite{huang2019coca,chartsias2017multimodal,liu2021unified,sharma2019missing,zhou2020hi}.

For the task of translating medical images, diffusion models inherently lack the ability to maintain the structural information accurately, which is due to the fact that the structured details of the source domain images are lost during the forward diffusion process and cannot be completely recovered through learned reverse diffusion, although it would be crucial to preserve the integrity of anatomical structures in medical images. To mitigate the aforementioned problem, Li et al. \cite{li2023zero} introduce a novel approach for structure-preserving image translation using frequency-domain filters, the Frequency-Guided Diffusion Model (FGDM). The proposed FGDM architecture enables zero-shot learning and can be exclusively trained on target domain data. Furthermore, their model does not require exposure to source domain data for training and can be directly deployed for source-to-target domain translation. The proposed method shows significant advantages in zero-shot medical image translation over the baseline SOTA methods.

\subsection{Reconstruction} \label{sec:reconstruction}
Medical image reconstruction plays a critical role in medical imaging. Its main goal is to produce high-quality medical images for clinical use while minimizing costs and patient risk \cite{levac2022accelerated,cao2022spirit}. Medical imaging modalities such as CT and MRI are medicine's most popular imaging tools. However, their physics restricts their efficacy, directly affects their performance, and degrades their desired results. The high resolution and complete result acquisition from the subject requires a higher radiation dose and a relatively longer resting time in the tube, which are only partially applicable due to the health precautions and the patient's relentless. Therefore, faster acquisition speed in medical imaging techniques like CT, Positron Emission Tomography (PET), and MRI is essential for reducing exam time, improving access to imaging services and reducing waiting times, and, more importantly, producing accurate images, particularly in dynamic studies that require fast imaging sequences. Accordingly, their radiation exposure reduces from the standard dose, or the imaging process is done in an under-sampled or sparse-view manner \cite{hyun2018deep,korkmaz2022unsupervised,feng2021task,korkmaz2021deep}. To diminish these drawbacks, e.g., low Signal-to-Noise Ratio (SNR) and Contrast-to-Noise Ratio (CNR), medical image reconstruction must overcome the challenges mentioned and solve this ill-posed inversion problem \cite{gothwal2022computational}. This section overviews the diffusion-based paradigms for medical image reconstruction and enhancement.

MRI is a popular non-invasive imaging utility in medical diagnosis treatment, but due to its innate physics, it is a time-consuming process of imaging sessions in which the movement of patients results in various artifacts in images. Therefore, to decrease bedtime and to accelerate the reverse problem-solving from the spatial domain (or \textit{k-space}) to image level, miscellaneous solutions are provided in the supervised-learning concept. However, these methods are not robust to distribution changes or drifts in their train/test sets. Jalal et al. \cite{jalal2021robust} proposed the first study in the MRI reconstruction domain via Compressed Sensing with Generative Models (CSGM). To this end, CSGM trains the score-based generative models \cite{song2020improved} on MRI images to utilize as prior information for the inversion pathway in reconstructing realistic MRI data from under-sampled MRI in posterior sampling scheme with Langevin dynamics \cite{bakry1985diffusions}. CSGM \cite{jalal2021robust} demonstrated its better performance over fastMRI \cite{zbontar2018fastmri} and Stanford MRI \cite{stanfordmri} datasets with SSIM and PSNR metrics in comparison with end-to-end supervised-learning paradigms.

\begin{figure}
	\centering
	\includegraphics[width=\columnwidth]{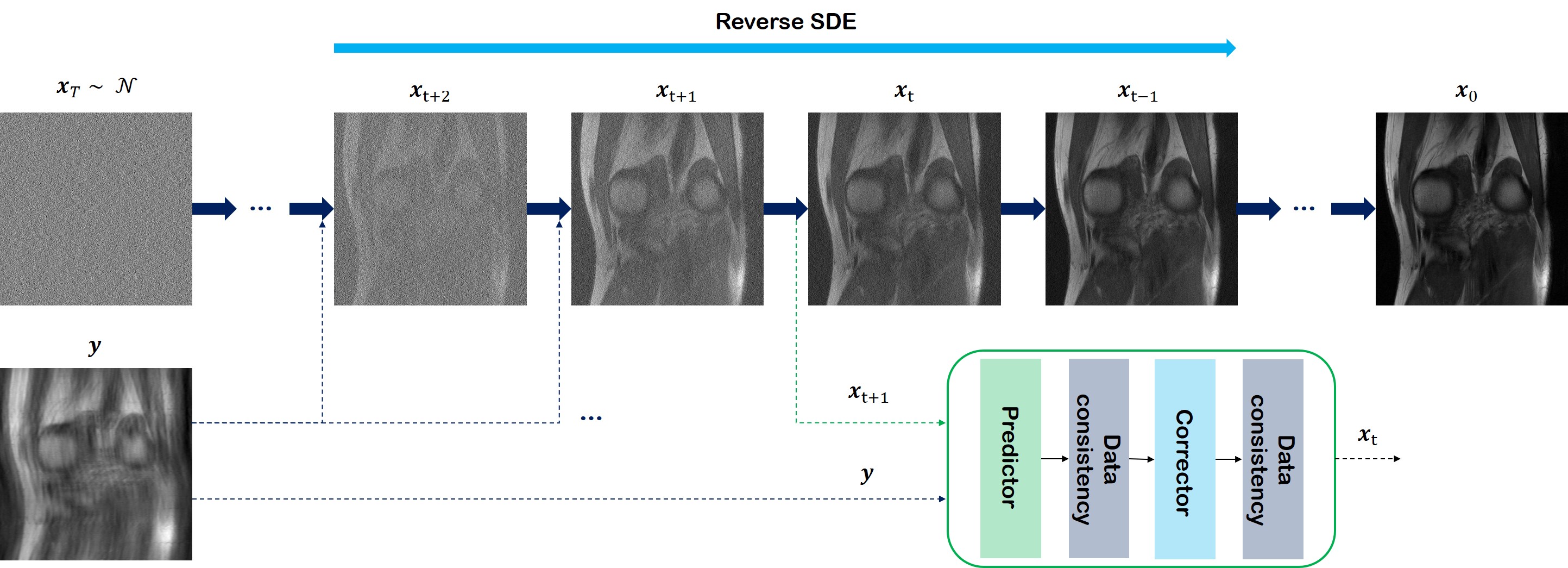}
	\caption{An overview of Score-MRI \textbf{\cite{chung2022score}}. $x_T$ is derived from the pre-trained prior distribution, and $x_0$ is retrieved by sequentially applying the predictor, data consistency, corrector, and data consistency steps, given the measurement through the reverse SDE procedure.}
	\label{fig:score-MRI}
\end{figure}

Chung et al. \cite{chung2022score} propose a score-based diffusion framework that solves the inverse problem for image reconstruction from accelerated MRI scans, as depicted in \Cref{fig:score-MRI}. In the first step, a single continuous time-dependent score function with denoising score matching is trained only with magnitude images. Then, in the reverse SDE process, the Variance Exploding (VE)-SDE \cite{song2020score} is exploited to sample from the pretrained score model distribution, conditioned on the measurement. 
Afterward, the image is first split into real and imaginary components at each step. Each part is fed into the predictor, followed by data consistency mapping to reconstruct the image. The obtained image is split again, and the correcter and the data consistency mapping are applied to each part to compensate for errors during the diffusion and reconstruct the improved image, respectively. Results demonstrate that the proposed model outperforms the previous SOTA methods \cite{block2007undersampled,zhou2020dudornet,zbontar2018fastmri} and can even reconstruct the data, which is considerably outside the training distribution with high fidelity, e.g., reconstructing anatomy not seen during training. In addition, the proposed framework has shown to be very effective for reconstructing the image when multiple coils exist. For the aforementioned problem, they present two approaches: (1) they reconstruct each coil image in parallel; (2) they take into account the correlation between the coil images by injecting the dependency between them at each given step during reverse SDE. Then, the final image is acquired by taking the sum-of-root-sum-of-squares (SSoS) \cite{roemer1990nmr} of each reconstructed coil image. Although these two techniques have shown great results  qualitatively and practically, they are time-consuming.

Liu et al. \cite{liu2022dolce} addressed the limited-angle CT reconstruction with a model-based DDPM paradigm called DOLCE. Based on the Fourier slice theorem, the conventional algorithm for mapping CT images from sinograms is Filtered Back Projection (FBP) \cite{kak2001principles}. Therefore, the limited-angle measurements can lead to the loss of Fourier measurements and, thus, degraded reconstruction results. However, due to the ill-posed nature of the reconstruction framework, the DDPM can not be utilized directly. DOLCE \cite{liu2022dolce} incorporates the output of FBP on the limited sinograms as prior information to condition the diffusion model. Further, due to the consistency conditions presented by sinograms, DOLCE imposes a consistency term within an additional step in the denoising iteration under $\ell_2$-norm loss in the inference step. The results of the Kidney CT (C4KC-KiTS) dataset \cite{heller2020state} regarding SSIM and PSNR metrics demonstrate that DOLCE is effective in producing sharp CT images.

\subsection{Registration} \label{sec:registration}
Deformable image registration is a critical medical image analysis technique focused on identifying non-rigid relationships between a pair of moving and fixed images. It plays a vital role when image shapes change due to factors such as subject, scanning time, and imaging modality. Traditional registration algorithms can be computationally expensive, while deep-learning methods are faster but still struggle with realistic continuous deformations.

To overcome these limitations, Kim et al. \cite{kim2022diffusemorph} introduce a novel diffusion-based method called DiffuseMorph. DiffuseMorph has two main networks - a diffusion network and a deformation network - both trained in an end-to-end fashion. The diffusion network scores the deformation between the moving and fixed images, while the deformation network uses this information to estimate the deformation field. The information used contains spatial information, allowing the generation of deformation fields along a continuous trajectory from the moving to the fixed image. The moving image is then transformed into a deformed image using the generated deformation fields and the spatial transformation layer (STL) \cite{jaderberg2015spatial}. In the inference phase, the model can provide both image registration and generation tasks. The experimental results affirm the high accuracy of the proposed method in registering both 2D facial expressions \cite{langner2010presentation} and 3D medical images \cite{lamontagne2019oasis,bernard2018deep}.

\subsection{Classification} 
\label{sec:classification}

The classification task is of great importance in medical image analysis, as it allows for accurately identifying and characterizing different structures and anomalies within medical images. Its ability to aid medical professionals in interpreting large amounts of complex data has the potential to revolutionize the healthcare industry \cite{shehab2022machine}. Despite this potential, adopting diffusion models to enhance classification results remains a significant challenge that needs to be addressed further.

\begin{figure}
	\centering
	\includegraphics[width=\columnwidth]{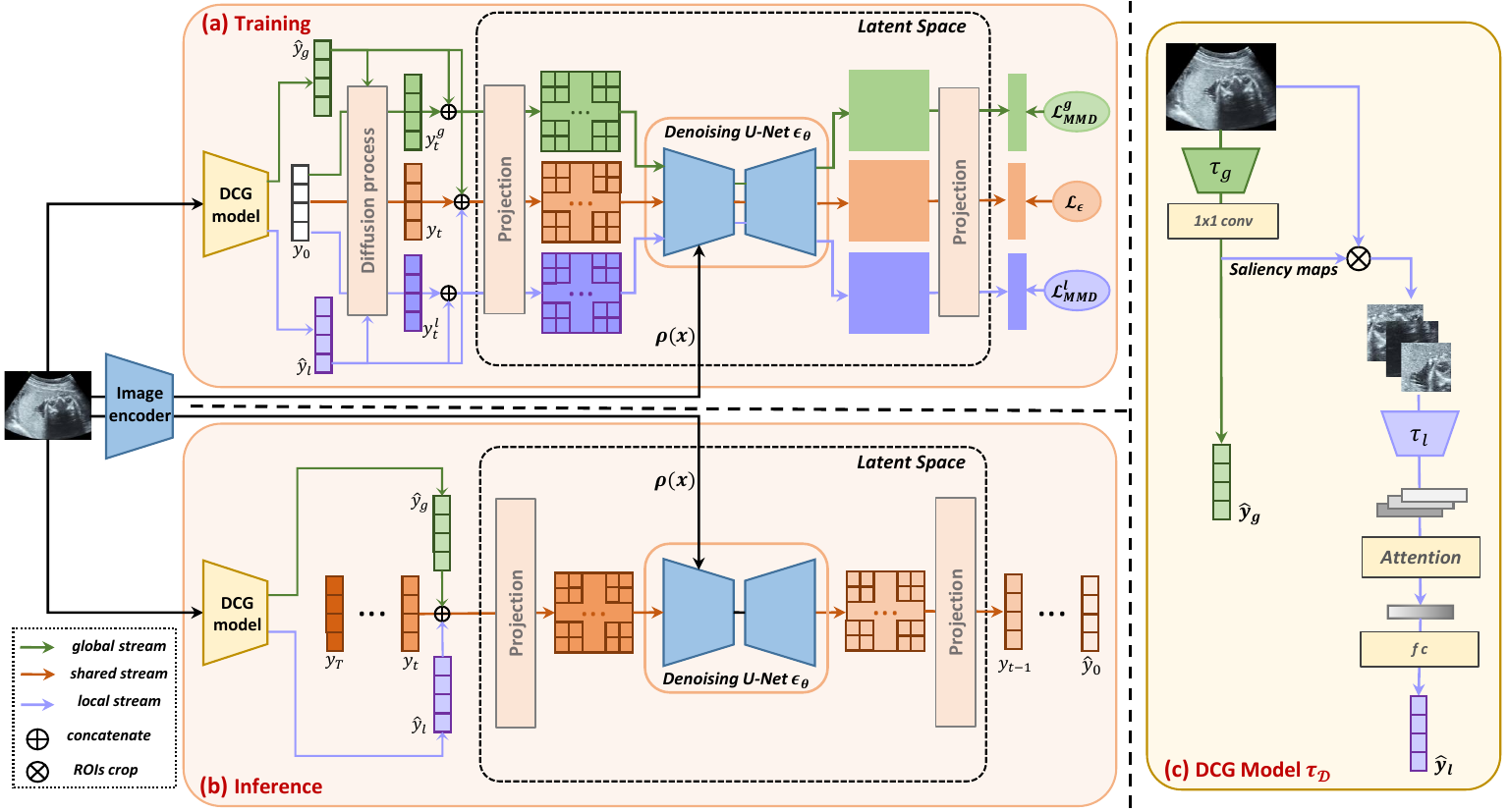}
	\caption{The image depicts the two stages of DiffMIC \cite{yang2023diffmic} - the training (a) and inference (b) stages - as well as the Dual-granularity Conditional Guidance (DCG) model shown in (c). The DCG model generates global and local priors from the raw image and Regions of Interest (ROIs), which serve to guide the diffusion process.}
	\label{fig:diffmic}
\end{figure}

DiffMIC \cite{yang2023diffmic} presents a new approach for classifying different medical image modalities using diffusion models, as shown in \Cref{fig:diffmic}. It first encodes the input image into a feature embedding space and uses a Dual-granularity Conditional Guidance (DCG) model to capture global and local prior information. The ground truth and two priors are then diffused to generate three noisy variables, which are concatenated with their corresponding priors and projected to a latent space to obtain three feature embeddings. The denoising U-Net integrates these embeddings with the image feature embedding and predicts the noise distribution for each embedding. Next, the feature embeddings obtained are projected back to their original dimensions. In order to estimate the amount of noise added to both the global and local priors, as well as the ground truth, DiffMIC utilizes maximum-mean discrepancy (MMD) \cite{gretton2006kernel,li2015generative} regularization loss and mean squared error (MSE) loss, respectively. In the inference stage, the DCG model is used to acquire dual priors from the input image, and the final prediction is denoised iteratively using the trained U-Net conditioned by the dual priors and the image feature embedding. Overall, DiffMIC presents a promising approach for accurately classifying medical images using diffusion models for the three considered tasks: placental maturity grading using ultrasound images, skin lesion classification using dermatoscopic images, and diabetic retinopathy grading using fundus images.

\subsection{Segmentation} \label{sec:segmentation}
Image segmentation is a vital task in computer vision, which investigates simplifying the complexity of the image by decomposing an image into multiple meaningful image segments \cite{azad2022contextual,heidari2023hiformer}. Specifically, it facilitates medical analysis by providing beneficial information about anatomy-related areas. However, deep learning models often require vast amounts of diverse pixel-annotated training data in order to produce generalizable results \cite{azad2022transnorm,chen2021transunet}. Nonetheless, the number of images and labels accessible for medical image segmentation is restricted due to the time, cost, and expertise required \cite{azad2022transdeeplab,cao2021swin,aghdam2022attention}. To this end, diffusion models have emerged as a promising approach in image segmentation research by synthesizing the labeled data and obviating the necessity for pixel-annotated data.

brainSPADE \cite{fernandez2022can} proposes a generative model for synthesizing labeled brain MRI images that can be used for training segmentation models. brainSPADE is composed of a label generator and an image generator sub-model. The former is responsible for creating synthetic segmentation maps, and the latter for synthesizing images based on generated labels. In the label generator, the input segmentation map is first encoded during training using a spatial VAE encoder and builds a latent space. The compressed latent code is then diffused and denoised via LDMs \cite{rombach2022high} and produces an efficient latent space in which imperceptible details are ignored, and semantic information is highlighted more. A spatial VAE decoder then constructs the artificial segmentation map via the latent space. In the image generator, Fernandez et al. \cite{fernandez2022can} take advantage of SPADE \cite{park2019semantic}, a VAE-GAN model, to build a style latent space from the input arbitrary style and use it with the artificial segmentation map to decode the output image. nnU-Net \cite{isensee2021nnu} was leveraged to examine the performance. Findings show that the model achieves comparable results when trained on synthetic data compared to that trained on factual data, and their combination significantly improves the model result. 

\begin{figure}
	\centering
	\includegraphics[width=\columnwidth]{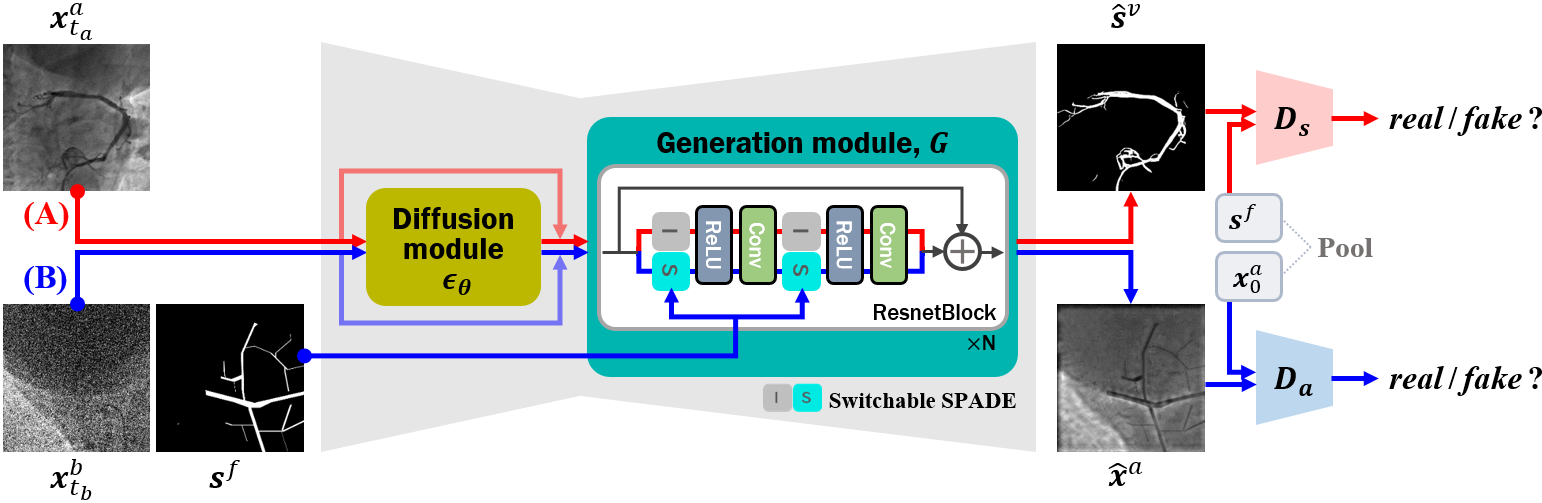}
	\caption{An overview of DARL \cite{kim2022vessel}. Path \textcolor{red}{(A)} involves feeding a real noisy angiography image through the model to generate a segmentation map. Path \textcolor{blue}{(B)} incorporates passing a noisy background image alongside a vessel-like fractal mask through the model to synthesize a synthetic angiography image.}
	\label{fig:DARL}
\end{figure}

Kim et al. \cite{kim2022vessel} propose a novel diffusion adversarial representation learning (DARL) model for self-supervised vessel segmentation, aiming to diagnose vascular diseases. There are two main modules in the proposed DARL model: a diffusion module, which learns background image distribution, and a generation module, which generates vessel segmentation masks or synthetic angiograms using a switchable SPADE algorithm \cite{park2019semantic}. \Cref{fig:DARL} illustrates two ways in which this method can be applied. In path (A), a real noisy angiography image $x^a_{t_{a}}$ is input into the model to produce a segmentation mask $\hat{s}^v$, and the SPADE switch is off. In path (B), a real noisy background image $x^b_{t_{b}}$ is fed into the model, and the SPADE becomes active and receives a vessel-like fractal mask, generating a synthetic angiography image $\hat{x}^a$. Then, by giving the generated synthetic angiography images into the path (A), a cycle is formed, which helps in learning the vessel information. In addition, during inference, path (A) is performed at one step, where the model produces the mask by only inputting the noisy angiography image into the model. Results verify the generalization, robustness, and superiority of the proposed method compared to SOTA un/self-supervised learning approaches.

In addition to the studies mentioned earlier, Rahman et al. \cite{rahman2023ambiguous} introduced the CIMD framework, a single probabilistic diffusion-based model, to address the ambiguous medical image segmentation task. Deterministic medical image segmentation frameworks such as \cite{aghdam2022attention,milletari2016v} produce a pixel-wise uncertainty, but the results are not consistent \cite{kendall2017uncertainties}. Also, from a medical aspect, medical image segmentation should not be considered just as a pixel-wise task. In clinical practice, analyzing organs or other structures from medical images is not a deterministic pixel-wise process but underlies the assessment of the whole image or, on a smaller scale, assessing the neighboring pixels' diversity.
The stochastic sampling step in the diffusion model can produce diverse and multiple masks. During its training step, the CIMD \cite{rahman2023ambiguous} utilizes the noisy segmentation ground-truth masks concatenated to the original image to hinder the conventional usage of the diffusion process in the segmentation task from producing more resilient results, rather than arbitrary masks. CIMD outperforms the probabilistic U-Net model \cite{kohl2018probabilistic} in terms of Collective Insight Score, which Rahman et al. \cite{rahman2023ambiguous} introduced in their work, which investigates three datasets (one private and two publicly available ones) with different modalities \cite{armato2004lung,carass2017longitudinal}.

Bieder et al. \cite{bieder2023diffusion} present a memory-efficient patch-based diffusion model called PatchDDM that can be applied to large 3D volumes, making it suitable for medical tasks. The authors evaluate PatchDDM on the tumor segmentation task of the BraTS2020 \cite{menze2014multimodal,bakas2018identifying} dataset and demonstrate that it can generate meaningful three-dimensional segmentation while requiring less computational resources than traditional diffusion models.

\subsection{Denoising} \label{sec:denoising}
The major challenge in medical imaging is obtaining an image without losing important information. The images obtained may be corrupted by noise or artifacts during the acquisition and/or further processing stages \cite{zhang2021transct,luthra2021eformer}. Noise reduces the image quality and is especially significant when the imaged objects are small and have relatively low contrast \cite{kazeminia2020gans}. Due to the nature of generative models, diffusion models are convenient for diverse denoising problems \cite{xia2022low,gao2023corediff}. In this section, we will explore the contribution of diffusion models to this task.

\begin{figure*}[!t]
	\centering
	\begin{subfigure}[][][c]{0.58\textwidth}
		\includegraphics[width=\textwidth]{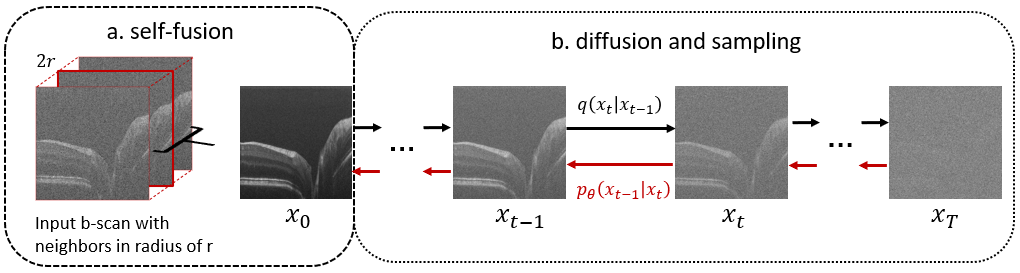}
		\caption{}
		\label{fig:DenoOCT-DDPM}
	\end{subfigure}
	\hfill
	\begin{subfigure}[][][c]{0.4\textwidth}
		\begin{tabular}{cccc}
			& \tiny \textbf{SNR=92dB} & \tiny \textbf{SNR=96dB} & \tiny \textbf{SNR=101dB} \\
			\rotatebox{90}{\hspace{0cm} \tiny \textbf{5-mean}} &
			{\includegraphics[width=0.25\linewidth]{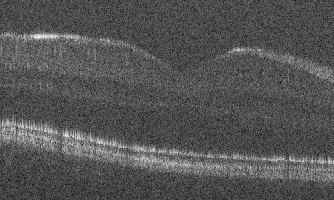}} &
			{\includegraphics[width=0.25\linewidth]{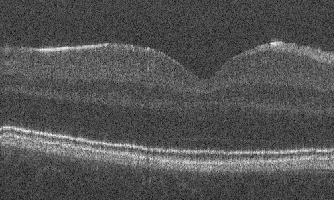}} &
			{\includegraphics[width=0.25\linewidth]{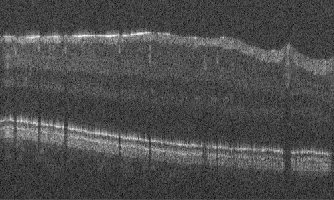}}  \\
			
			\rotatebox{90}{\hspace{0cm} \tiny \textbf{PMFN}}&
			{\includegraphics[width=0.25\linewidth]{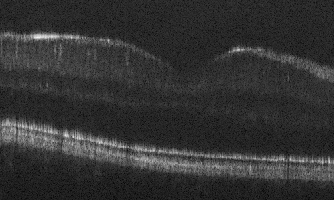}} &
			{\includegraphics[width=0.25\linewidth]{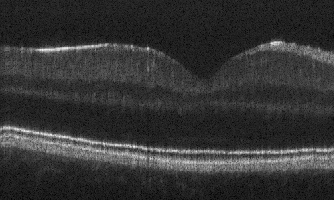}} &
			{\includegraphics[width=0.25\linewidth]{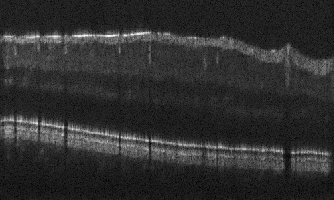}}  \\
			
			\rotatebox{90}{\hspace{0cm} \tiny \textbf{Method}} &
			{\includegraphics[width=0.25\linewidth]{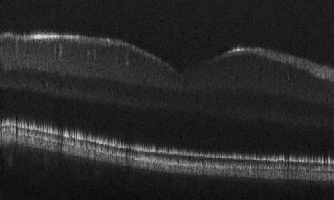}} &
			{\includegraphics[width=0.25\linewidth]{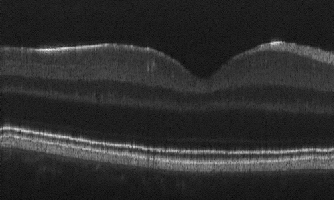}} &
			{\includegraphics[width=0.25\linewidth]{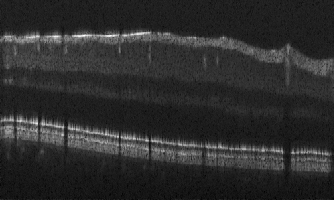}}  \\
		\end{tabular}
		\caption{}
		\label{fig:DenoOCT-DDPM-results}
	\end{subfigure}
	\caption{(Left) General pipeline of DenoOCT-DDPM \cite{hu2022unsupervised}. To feed the diffusion model more low-noise reference images, DenoOCT-DDPM applies a self-fusion \cite{oguz2020self} on the \textcolor{red}{red} canvas b-scan in (a) with neighboring b-scans for higher SNR b-scan. (b) indicates a straightforward DDPM scheme in an unsupervised manner to learn speckle noise distribution in the red arrow sampling stream. (Right) Ultimate visual comparison of DenoOCT-DDPM \cite{hu2022unsupervised} for denoising b-scans of volumetric OCT images demolished by speckle noise. DenoOCT-DDPM compared itself with PMFN \cite{hu2020retinal} and the average of 5 successive b-scans as the ground truth. PMFN results in good feature preservation in diverse retinal layers, but it can easily over-smooth small vessel regions. In contrast, the DenoOCT-DDPM retinal layers are more homogenous than PMFN in diverse acquisition SNRs.}
	\label{fig:DenoOCT-DDPM-all}
\end{figure*}

Hu et al. \cite{hu2022unsupervised} utilized a DDPM \cite{ho2020denoising} to despeckle Optical Coherence Tomography (OCT) volumetric retina data in an unsupervised manner, denoted as DenoOCT-DDPM. OCT imaging utility suffers from limited spatial-frequency bandwidth, which leads to the resulting images containing speckle noise. Speckle noise hinders the ophthalmologist's diagnosis and can severely affect the visibility of the tissue. The classic methods, such as averaging multiple b-scans at the same location, have extreme drawbacks, such as prolonged acquisition time and registration artifacts. Due to the multiplicative properties of speckle noise, these methods enrich rather than reduce the noise. Deep-based models perform outstandingly. This performance, however, depends on the availability of noise-free images, which is a rare and costly process to acquire. To this end, DenoOCT-DDPM \cite{hu2022unsupervised} utilizes DDPM's feasibility in noise patterns rather than real-data pattern. Therefore, they use a self-fusion \cite{oguz2020self} as a preprocessing step to feed the DDPM with a clear reference image and train the parameterized Markov chain (see \Cref{fig:DenoOCT-DDPM}). Their investigation demonstrated the SOTA results over the Pseudo-Modality Fusion Network (PMFN) \cite{hu2020retinal}, which uses information from the single-frame noisy b-scan and a pseudo-modality that is created with the aid of the self-fusion \cite{oguz2020self} method, regarding Signal-to-Noise Ratio (SNR) metric. The qualitative results over PMNF depicted in \Cref{fig:DenoOCT-DDPM-results} (represented in multiple acquisition SNRs) endorse the ability of diffusion models to remove the speckle noise while preserving the fine-grained features like small vessels.

\begin{figure*}[t]
	\centering
	\includegraphics[width=0.95\textwidth]{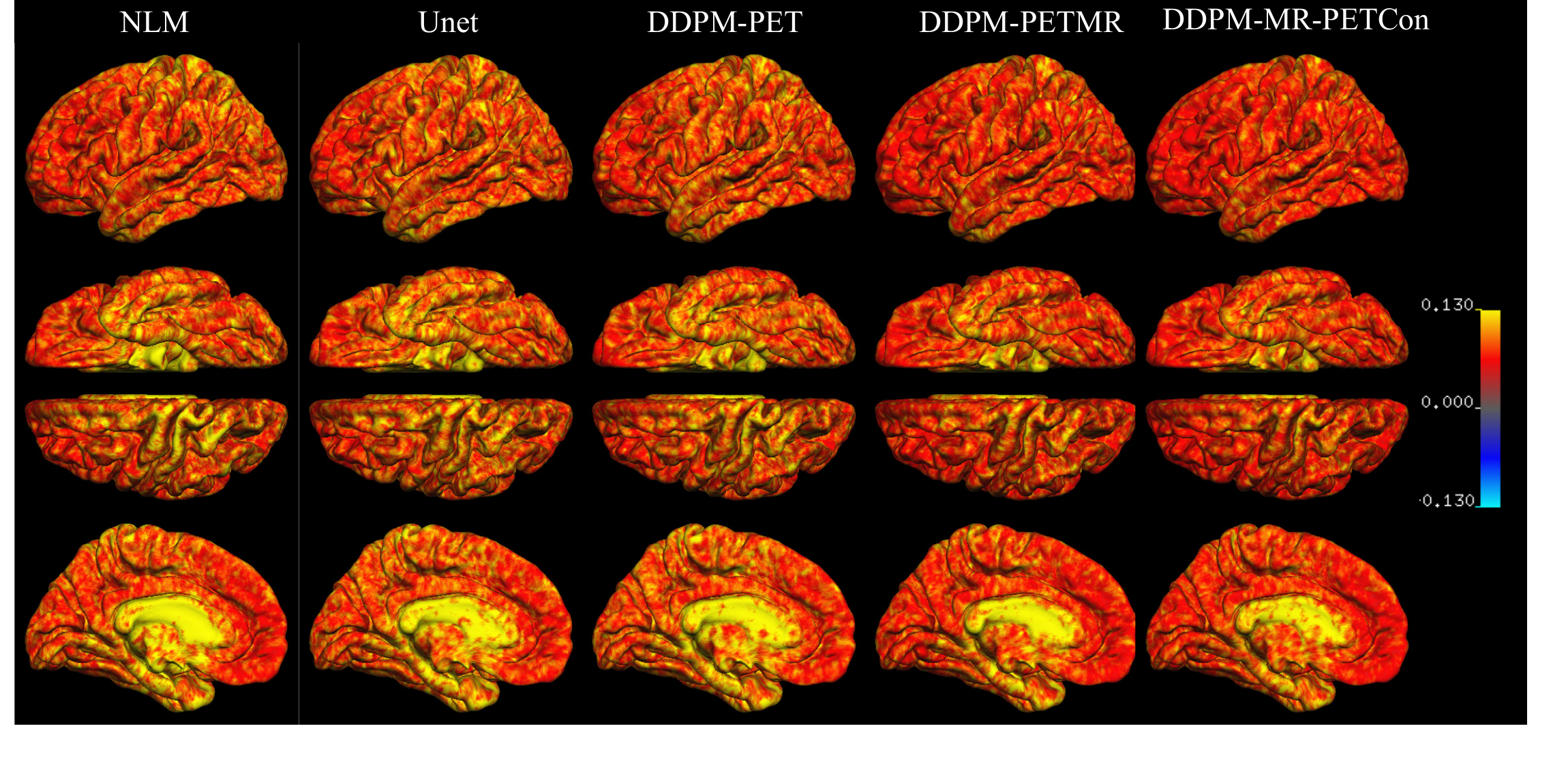}
	\caption{Comparison by different methods of surface error mapping of the left hemisphere from 20 $^{18}$F-MK-6240 test dataset \cite{gong2022pet}. The results confirm that DDPM-MR-PETCon has the lowest error, followed by the DDPM-PETMR method. DDPM-MR-PETCon is short for using the MR image as network input while using the PET image as a data consistency item, and DDPM-PETMR denotes that PET and MR images are used as network input.}
	\label{fig:PET-DDPM}
\end{figure*}

PET is a non-invasive imaging utility that plays a crucial role in cancer screening and diagnosis. However, as with OCT devices, PETs suffer from low SNR and resolution due to the low beam count radiation to patients. Deep learning methods over PET image denoising have advanced, but over-smoothing is a prominent drawback of CNN-based approaches. Therefore, Conditional Generative Adversarial Networks (CGANs) \cite{zhou2020supervised,song2020pet} neutralize the mentioned deficiency but still depend on training and test set distributions. Gong et al. \cite{gong2022pet} proposed the DDPM-based framework for PET denoising in collaboration with an assistive modality embedding as prior information to DDPM formulation, namely PET-DDPM. Gong et al. used $^{18}$F-FDG and  $^{18}$F-MK-6240 datasets for PET and MR modalities, respectively. PET-DDM is a multi-disciplinary study investigating the excessive modalities of collaboration in learning noise distribution through PET images. This intuition follows the original paper in a generative paradigm \cite{dhariwal2021diffusion} with a guided classifier to converge the learned distribution to desired distribution. As qualitatively illustrated in \Cref{fig:PET-DDPM}, PET-DDM produced SOTA results compared with U-Net \cite{ronneberger2015u} based denoising network in terms of PSNR and SSIM.

Diffusion MRI is an important modality for studying oncologic and neurologic biomarkers but suffers from long acquisition times and low SNR. Xiang et al. \cite{xiang2023ddm} explore these deficiencies by imposing the self-supervised statistic-based denoising strategy into diffusion models and by performing denoising through the conditional generation process. To this end, their approach consists of three main stages. First, they learn an initial noise distribution in a self-supervised manner. Next, they estimate a noise model with $\mu=0$ and $\sigma$ (a Gaussian distribution) from the learned noise distribution in the previous step. They apply a $p-$norm to minimize the distance between the $\sigma$ and diffusion sampling noise $\beta$. Ultimately, they train another diffusion model to produce clean images in an unsupervised manner. They elaborated their findings on one private and three public datasets (Sherbrooke 3-Shell \cite{garyfallidis2014dipy}, Stanford HARDI \cite{Rokem2016}, and Parkinson’s Progression Markers Initiative (PPMI) \cite{ppmi}) and reported superior denoising performances.

\subsection{Image Generation}\label{sec:generation}
Image generation is one of the primary objectives of diffusion models, which has been widely applied in a variety of styles, including generating synthetic 2D/3D medical images \cite{pinaya2022brain,moghadam2023morphology,dorjsembe2022three,kim2022diffusion}, reconstructing 3D cell from 2D cell images \cite{waibel2022diffusion}, etc. This section will outline the diffusion-based approaches for medical image generation.

\begin{figure*}[t]
	\centering
	\begin{subfigure}{0.49\textwidth}
		\includegraphics[width=\textwidth]{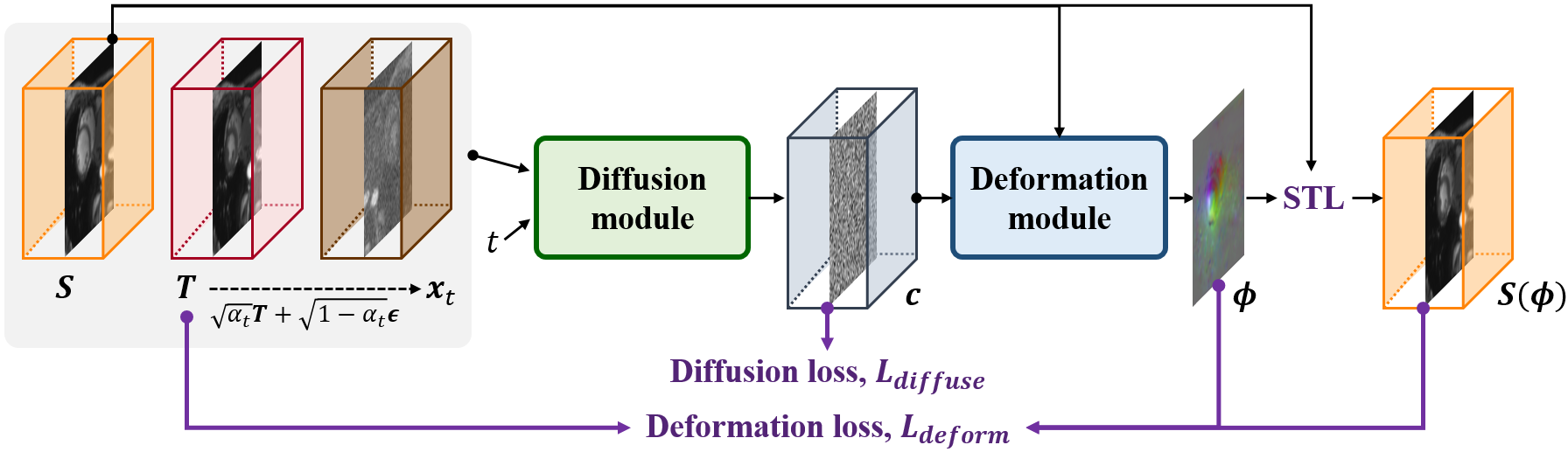}
		\caption{}
		\label{fig:train-DDM}
	\end{subfigure}
	\begin{subfigure}{0.49\textwidth}
		\includegraphics[width=\textwidth]{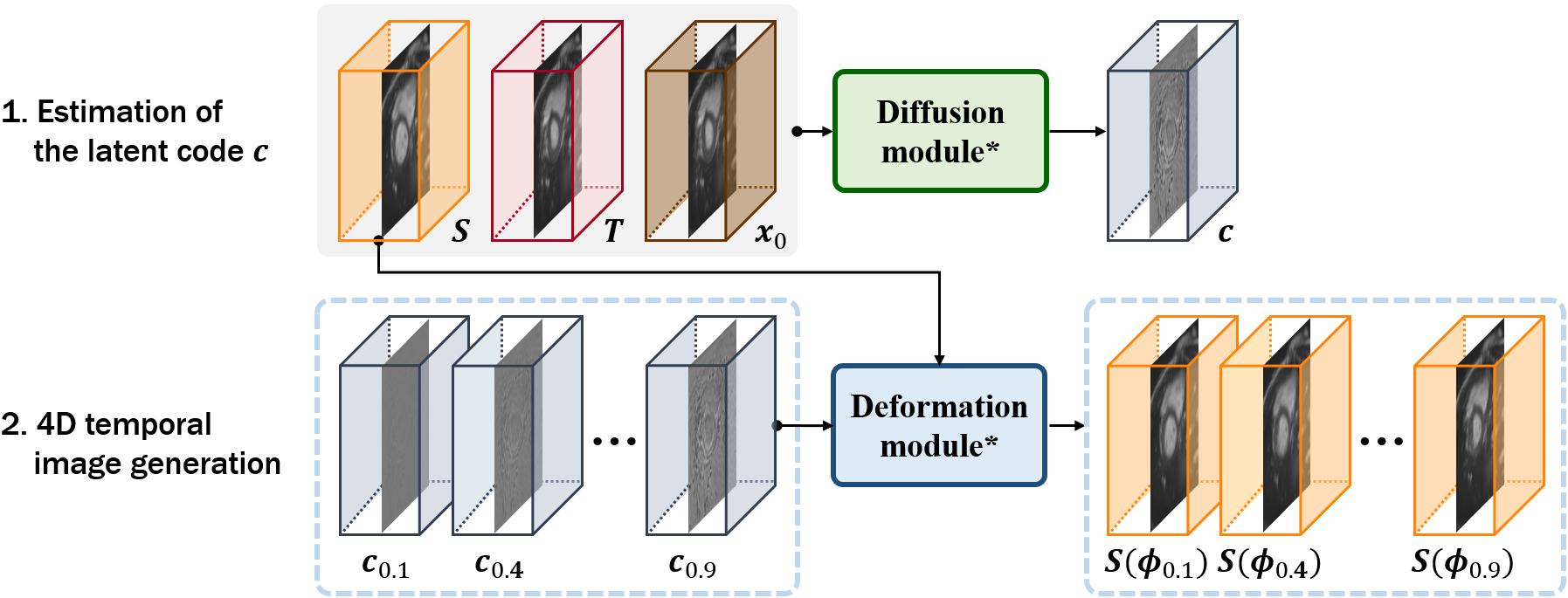}
		\caption{}
		\label{fig:inference-DDM}
	\end{subfigure}
	\caption{\textbf{(a)} demonstrates the DDM  \cite{kim2022diffusion} training phase and \textbf{(b)} the inference phase.}
\end{figure*}

\begin{figure*}[!t]
	\centering
	\includegraphics[width=0.85\textwidth]{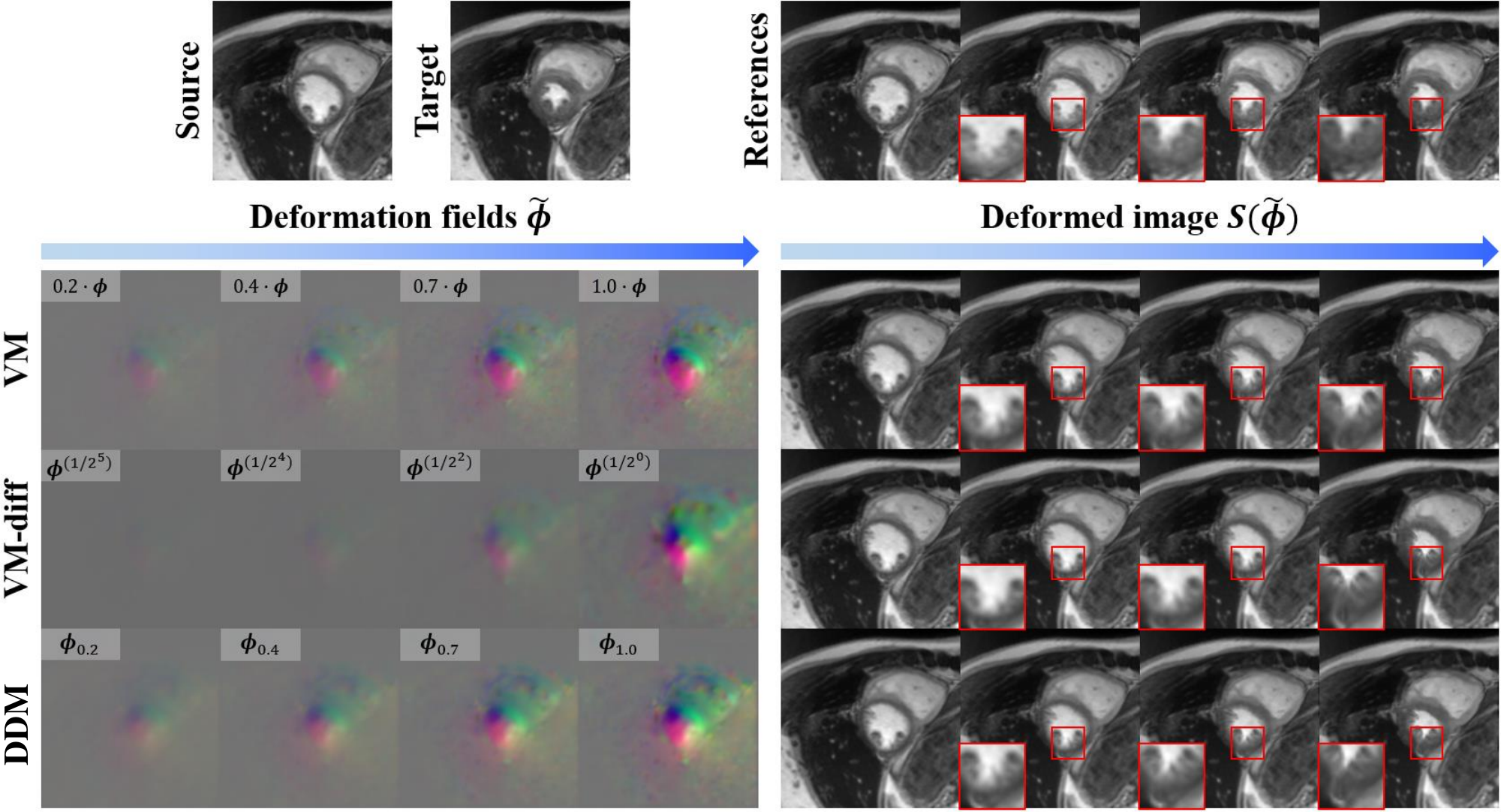}
	\caption{Visual comparison of DDM \cite{kim2022diffusion}, VM \cite{balakrishnan2018unsupervised}, and VM-Diff \cite{dalca2018unsupervised} for generating temporal cardiac images. The deformed intermediate frames $S(\tilde{\phi})$ (right) are constructed using the source and target, and produce the deformation fields $\tilde{\phi}$ (left).}
	\label{fig:DDM-result}
\end{figure*}

Using 4D imaging to follow anatomical changes is one of the methods used in medicine to track 3D volumes over time to detect anomalies and disease progression. Such 4D images are primarily obtained with MRI, but this process is relatively time-consuming. Kim et al. \cite{kim2022diffusion} recently proposed the Diffusion Deformable Model (DDM), which takes source and target images and generates intermediate temporal frames along the continuous trajectory. This approach comprises two main modules: (i) a denoising diffusion probabilistic model (DDPM) module and (ii) a deformation module. In the DDPM module, a latent code is constructed by learning the source and target images, and in the deformation module, the acquired latent code and the source image are used to render the deformed image. In the training phase, as shown in \Cref{fig:train-DDM}, the diffusion model, derived from \cite{ho2020denoising}, takes source, target, and perturbed target images and outputs a latent code. The learned latent code along the source image is fed into the deformation module, adopted from \cite{balakrishnan2018unsupervised}, and creates deformation fields. Then, the spatial transformation layer (STL) \cite{jaderberg2015spatial}, with tri-linear interpolation, is employed to warp the source volume using the deformation fields in order to build the deformed source image. Afterward, inference begins with the diffusion module providing the latent code, which contains spatial information from the source toward the target (see \Cref{fig:inference-DDM}). Then, deformed intermediate frames are generated using the deformation module by scaling the latent code with a factor, which is an element of [0, 1]. Additionally, qualitative results are illustrated in \Cref{fig:DDM-result}.

Packhauser et al. \cite{packhauser2022generation} utilize a latent diffusion model \cite{rombach2022high} to produce high-quality class-conditional chest X-ray images while proposing a sampling strategy to maintain sensitive biometric information's privacy during the generation process. To assess the potential utility of the generated dataset, the images are evaluated on a thoracic abnormality classification task, and the results show that the proposed approach outperforms GAN-based methods.

Histopathology involves the study of tissues and cells at the microscopic level in order to diagnose diseases and cancer \cite{di2016feature}. Histology images, however, are rare for some cancer subtypes, thereby increasing the significance of generative models to fill the void. To this end, Moghadam et al. \cite{moghadam2023morphology} investigate adopting DDPMs for generating histopathology images for the first time. Specifically, they exploit the DDPMs with genotype guidance to synthesize images containing various morphological and genomic information. To tackle this data consistency problem and enforce the model to focus more on morphological patterns, they first feed input images into a color normalization module \cite{vahadane2016structure} to unify the domain of all images. In addition, they apply a morphology levels prioritization module \cite{choi2022perception} that designates higher weight values to the loss at earlier levels to emphasize perceptual information and lower weights to the loss at later levels, resulting in higher fidelity samples. Experiments on the Cancer Genome Atlas (TCGA) dataset \cite{grossman2016toward} exhibit the superiority of the proposed method compared to GAN-based approaches \cite{karras2017progressive}.

In the case of diffusion-based MRI synthesis models, it is common to employ a unimodal approach. However, since they rely on the original image domain, these models often suffer from high memory demands and are less practical for multi-modal synthesis purposes. To mitigate this problem, Jiang et al. \cite{jiang2023cola} propose the first diffusion-based multi-modality MRI synthesis model, namely the Conditioned Latent Diffusion Model (CoLa-Diff). Specifically, they propose an architecture designed to reduce memory consumption by operating in the latent space. In order to address potential issues with compression and noise present in the latent space, they utilize a cooperative filtering approach inspired by collaborative filtering techniques. Moreover, to ensure the preservation of anatomical structures, they consider the inclusion of brain region masks as priors for density distributions to guide the diffusion process. Additionally, they implement an auto-weight adaptation technique to leverage multi-modal information effectively. Their experiments demonstrate that the proposed method outperforms other SOTA MRI synthesis methods, indicating that CoLa-Diff has significant promise as an effective tool for facilitating multi-modal MRI synthesis.


\subsection{Anomaly Detection}
\label{sec:anomaly}
Medical anomaly detection is an important topic in computer vision, aiming to highlight the anomalous regions of the image \cite{chen2022utrad,tschuchnig2022anomaly,fernando2021deep,bercea2023reversing,shi2023dissolving}. Generative models have dramatically shaped queries on anomaly detection in recent years and have shown promising results. Accordingly, we explore diffusion models as dominant generative models in anomaly detection in the following section.

Wolleb et al. \cite{wolleb2022diffusionanomaly} introduce a weakly supervised learning method based on Denoising Diffusion Implicit Models (DDIMs) \cite{song2021denoising} for medical anomaly detection. Given an input image of a healthy or diseased subject, image-to-image translation first performs such that the objective is to translate the input image into the healthy one. Then, the anomaly regions are identified by subtracting the output image from the input. This process begins by encoding an input image into a noisy image with reversed DDIM sampling. Then, the denoising process is guided through a binary classifier trained beforehand on the healthy and diseased images to produce the healthy image. Finally, the anomaly map is calculated by taking the difference between the output and input. Results on BraTS2020 \cite{bakas2017advancing,bakas2018identifying,menze2014multimodal} and CheXpert \cite{irvin2019chexpert} datasets demonstrate the superiority of the proposed approach compared to both VAE \cite{zimmerer2018context} and GAN \cite{siddiquee2019learning} models.

Wyatt et al. \cite{wyatt2022anoddpm} in AnoDDPM train a DDPM only on healthy medical samples. The anomaly image is then rendered by computing the difference between the output and input images. They also show that leveraging Simplex noise over Gaussian noise significantly enhances the performance. 

\begin{figure}[!th]
	\centering
	\includegraphics[width=\columnwidth]{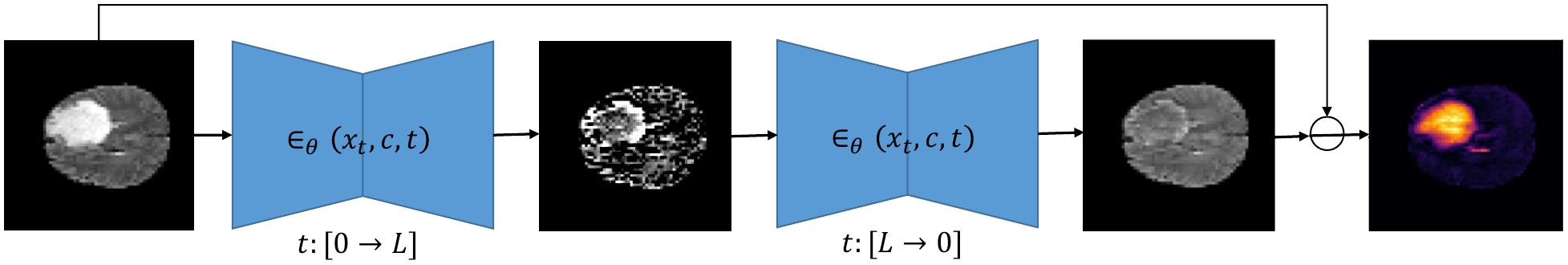}
	\caption{An overview of CDPM \cite{sanchez2022healthy}. Iteratively applying diffusion models using an unconditional model $(c=\emptyset)$ encodes the input image into a latent space. Then, reversing the diffusion process from the latent space decodes a healthy state image. The decoding process is guided by conditioning it on the healthy state and $\emptyset$. The anomaly heatmap is generated by subtracting the input image from the generated counterfactual.}
	\label{fig:CDPM}
\end{figure}

In contrast, CDPM \cite{sanchez2022healthy} demonstrates that training the diffusion probabilistic models only on healthy data generates poor segmentation performance. Thus, CDPM presents a counterfactual diffusion probabilistic model for generating healthy counterfactuals from factual input images. As illustrated in \Cref{fig:CDPM}, the input image is initially encoded into a latent space by iteratively applying diffusion models using an unconditional model. Then, the decoding step is accomplished by reversing the diffusion process. Using implicit guidance \cite{ho2021classifierfree}, the latent is decoded into a counterfactual by conditioning it on a healthy state and $\emptyset$. Inspired by \cite{ramesh2022hierarchical,nichol2021glide,rombach2022high}, 
Sanchez et al. \cite{sanchez2022healthy} then enhance the conditioning process by incorporating conditional attention into the U-Net backbone. As a final step, a dynamic normalization technique is applied during inference to avoid saturation in latent space pixels, caused by the guided iterative process that may change the image statistics. Eventually, the location of abnormality is determined by subtracting the input image from the generated healthy counterfactual.

\begin{figure}[h]
	\centering
	\includegraphics[width=0.85\columnwidth]{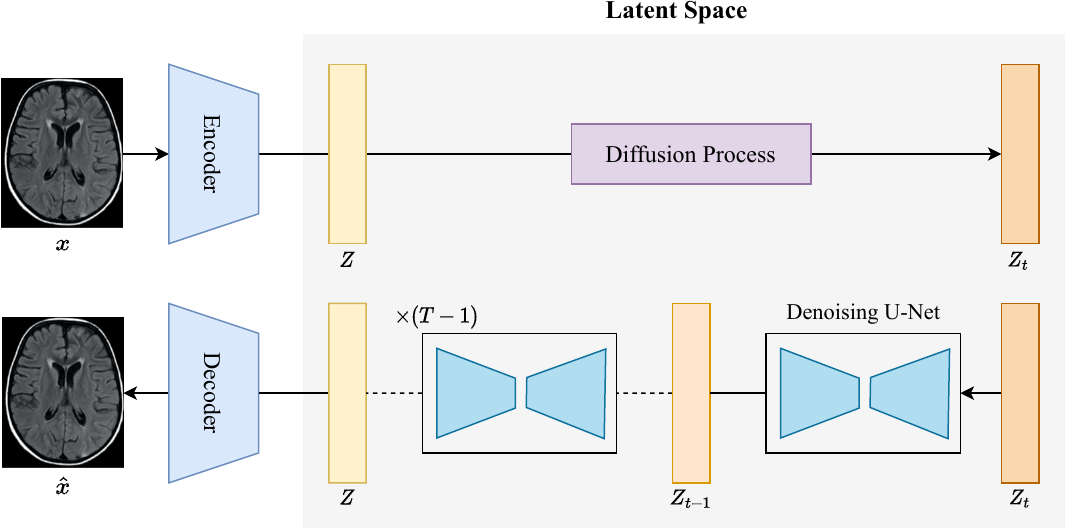}
	\caption{An overview of BAnoDDPM \cite{pinaya2022fast}. An autoencoder compresses the input image into a latent code, further enhanced by applying diffusion and reverse processes, and decodes into the pixel space.}
	\label{fig:LDM-method}
\end{figure}

Pinaya et al. \cite{pinaya2022fast} propose a fast DDPM-based approach for detecting and segmenting anomalous regions in the brain MR images (see \Cref{fig:LDM-method}). This method follows the strategy of generating a healthy sample and delineating the anomaly segmentation map by subtracting it from the input image. To this end, VQ-VAE \cite{van2017neural} is first adopted following \cite{rombach2022high}, which encodes the input images into a compact latent representation and provides the quantized latent representation from input images utilizing a codebook. The DDPM then uses the acquired latent space and learns the distribution of the latent representation of the healthy samples. A binary mask indicating the location of the anomaly is constructed by applying a pre-calculated threshold on the average of intermediate samples of the reverse process, which contain less noisy and more distinct values. Using the middle step as the starting point for the reverse process, they denoise the anomalous areas of the image and preserve the rest using the obtained mask, thereby removing the lesion from the sample. Eventually, upon decoding the sample, a healthy image is produced.

In recent work by Behrendt et al. \cite{behrendt2023patched}, the generation task of diffusion models is reframed as an estimation of healthy brain anatomy based on patches, utilizing spatial context to guide and enhance the reconstruction process. Specifically, they demonstrate that applying noise to the entire image simultaneously can pose challenges in achieving accurate reconstruction of the intricate structure of the brain. Therefore, patch-based Denoising Diffusion Probabilistic Models (pDDPMs) are introduced for Unsupervised Anomaly Detection (UAD) in brain MRI. In the proposed pDDPMs, they perform the forward diffusion process on a localized patch of the input image while employing the entire, partially noised image in the backward diffusion process to recover the noised patch. During inference, their trained pDDPM sequentially operates on sliding patches of the input image, alternately applying noise and denoising operations before stitching together the resultant denoised patches to yield the final denoised output image. Experiments on the public BraTS21 \cite{baid2021rsna} and MSLUB \cite{lesjak2018novel} datasets verify that their approach performs superior or on par with most contemporary works in terms of UAD performance.


\subsection{Other Applications and Multiple Tasks}
\label{sec:multi-task}
Based on \Cref{fig:taxonomy}, there are still studies that could not be assigned to a particular category, and the use of diffusion is not limited to those nine categories. Gong et al. \cite{gong2023diffusion} present an innovative semi-supervised learning framework that utilizes diffusion models to accurately quantify the brain midline shift observed in head CT images. Keicher et al. \cite{keicher2023semantic} introduced a new method for grading vertebral fractures using a Diffusion Autoencoder (DAE) as an unsupervised, generative feature extractor. Also, diffusion models are not limited to vision-related tasks in the medical domain and can also foster research innovations in biology; e.g. the platforms presented in \cite{trippe2023diffusion, anand2022protein} can be used for designing drugs and vaccines. In the following paragraphs, we explore some of the recent diffusion-based approaches used in multi-task learning and their distinctive uses in the medical domain.

Un/self-supervised learning is an ideal alternate approach in medical image denoising, where accessing paired clean and noisy images is difficult to achieve \cite{lehtinen2018noise2noise,kim2021noise2score}. Conventional deep-based networks utilize Minimum Mean Square Error (MMSE) estimates, which lead to unsatisfactory and blurred images due to the distribution change in train/test data or the preliminary assumption of Gaussian noise is at odds with the data's actual distribution. Chung et al. \cite{chung2022mr} proposed a multi-successive paradigm for MRI image denoising and super-resolution, namely R2D2+, with the SDE \cite{song2020score} algorithm to tackle the mentioned deficiencies. Diffusion generative models are robust to any distribution change over the data and produce more realistic data \cite{dhariwal2021diffusion}. Despite the advantages of diffusion models, they are very time-consuming. To this end, Chung et al. \cite{chung2022mr} do not start the reverse diffusion process from the pure noise but start from the initial noisy image. R2D2+ \cite{chung2022mr} solves a reverse time SDE procedure with a non-parametric estimation method based on eigenvalue analysis of covariance matrix rather than the conventional numerical methods \cite{song2020score}. To restrain structure alleviation through the process, R2D2+ uses a low-frequency regularization to hamper any change in the low-frequency counterparts of the image. R2D2+ utilizes the same network for the super-resolution task after the denoising step. The overall results over the single coiled fastMRI \cite{zbontar2018fastmri} knee dataset and private liver MRI dataset indicate the superiority of this approach over conventional SOTA un/self-supervised learning schemes in terms of SNR and Contrast-to-Noise Ratio (CNR) metrics.

Accurate models for diagnosing skin cancer are crucial for early detection and treatment. Current computer-aided systems use deep learning, but recent research has shown that these models are highly vulnerable to attacks that subtly alter images, causing them to misclassify skin lesions. To address this problem, a new defense method \cite{wang2022fight} is proposed that can reverse these distortions by using a multiscale image pyramid and injecting Gaussian noise at each scale to neutralize the effects of adversarial perturbations. A denoising mechanism is then employed to remove added noise and aggregate information from neighboring scales. By repeating this process, images become resistant to noise and achieve actual probability value. The final step involves fusing sub-images at different scales to produce a reversed image. Experimental results on the ISIC 2019 dataset \cite{combalia2019bcn20000} demonstrate the superiority of the proposed method in defending against different attacks.

\begin{figure}[!t]
	\centering
	\includegraphics[width=0.9\columnwidth]{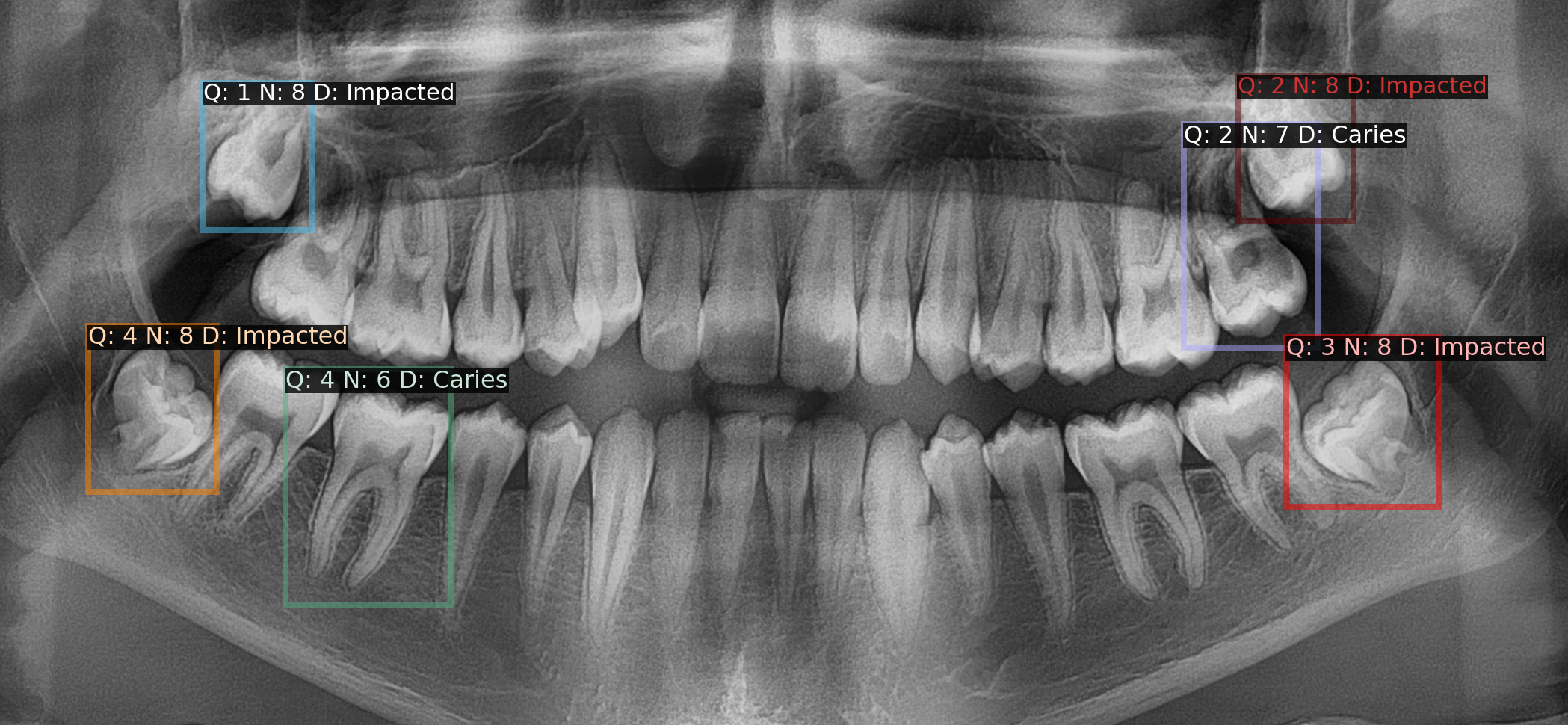}
	\caption{The final result of the HierarchicalDet \cite{hamamci2023diffusion} model displays bounding boxes around unhealthy teeth, along with the predicted quadrant (Q), enumeration (N), and diagnosis (D) labels.}
	\label{fig:diffusion-detection}
\end{figure}
Hamamci et al. \cite{hamamci2023diffusion} propose a novel method for detecting abnormal teeth in panoramic X-rays using a hierarchical multi-label approach. They employ the DiffusionDet model \cite{chen2022diffusiondet}, which utilizes a denoising diffusion process to predict objects and their categories from noisy boxes. In the first stage of their approach, the model is trained to predict quadrants and their corresponding bounding box coordinates. The researchers input a raw image into the encoder to create high-level features, which are then used in the decoder's denoising step to refine the bounding boxes. The second stage involves predicting the tooth number, quadrant number, and bounding box. However, instead of refining the complete noisy boxes, the researchers propose manipulating bounding boxes by concatenating inferred boxes from the previous stage with the noisy boxes. The encoder and decoder weights are transferred from the previous stage to this stage. Finally, the third stage uses a similar approach to detect abnormal teeth with their quadrant-enumeration-diagnosis label. The output of the final model is illustrated in \Cref{fig:diffusion-detection}. This hierarchical approach can handle partially labeled data and capture the full complexity of the underlying data. Additionally, the researchers present a new public dataset with three distinct data types: 1) for quadrant detection, 2) for tooth detection with both quadrant and tooth enumeration classifications, and 3) for diseased tooth detection with quadrant, tooth enumeration, and diagnosis classifications.

\subsection{Comparative Overview}
\label{overview}

\Cref{tab:paperhighlights} comprehensively categorizes the reviewed diffusion model papers according to which algorithm they directly used or inspired: (1) DDPMs, (2) NCSNs, and (3) SDEs. In addition, \Cref{tab:paperhighlights} highlights the key concepts and objectives of each algorithm and represents the practical use cases that can be investigated and utilized in future research based on reviewed papers. 

It is evident that conditioning the reverse diffusion process is one of the most studied methods for obtaining the desired output. This guiding process can be done using different constraint types. In \cite{lyu2022conversion,xie2022measurement,waibel2022diffusion}, they control the reverse process by applying conditions using images; in particular, Lyu et al. \cite{lyu2022conversion} condition the DDPM and SDE utilizing T2w MR images to obtain CT images, Xie et al. \cite{xie2022measurement} propose measurement-conditioned DDPM constituted for under-sampled medical image reconstruction, and Waibel et al. \cite{waibel2022diffusion} constrain the 3D model using 2D microscopy images for generating 3D single cell shapes. Moreover, BrainGen \cite{dar2022adaptive} produces realistic examples of brain scans, which are conditioned on meta-data such as age, gender, ventricular volume, and brain volume relative to intracranial volume. In addition, the use of a classifier and implicit guidance methods in \cite{wolleb2022diffusionanomaly} and \cite{sanchez2022healthy} have been investigated thoroughly. In this way, the distribution is shifted in a manner that is more likely to reach the expected outcome.

Some of the primary concerns and limitations of diffusion models are their slow speed and required computational cost. Several methods have been developed to address these drawbacks. Training-free Denoising Diffusion Implicit Model (DDIM) \cite{song2021denoising} is one of the advancements designed to accelerate the sampling process. DDIM extends the DDPM by substituting the Markovian process with the non-Markovian one, resulting in a faster sampling procedure with negligible quality degradation. \cite{cao2022high,chung2022come} propose a suitable initialization instead of a random Gaussian noise in the reverse process, causing a significant acceleration. Specifically, Chung et al. \cite{chung2022come} prove that, based on stochastic contraction theory, beginning the reversion, for example, after one step prediction of the pre-trained neural network, fastens the reverse diffusion and reduces the number of reverse samplings. Dar et al.  \cite{dar2022adaptive} also verify that adversarial learning can boost reverse diffusion speed by two orders of magnitude.

Several methods have also been worked on to enhance the output quality of diffusion models. AnoDDPM \cite{wyatt2022anoddpm} substantiates that generalization to other types of noise distributions can enhance task-specific quality. They ascertain that in the case of anomaly detection, Simplex noise shows superiority over Gaussian noise. Furthermore, Cao et al. \cite{cao2022high} corroborate that operating the diffusion process only in the high-frequency part of the image improves the stability and quality of the MRI reconstruction.

Despite the mentioned improvements in diffusion models, there remains a need to investigate why diffusion models have become popular in medical imaging and why some tasks are more successful in adopting diffusion models. Primarily, diffusion models have been increasingly used in medical imaging due to their effectiveness, ease of implementation, and high quality of output. In medical imaging, it is crucial to have high-resolution images that provide accurate local information for disease detection. Diffusion models have been able to achieve this, leading to their growing popularity in this field.

\begin{figure*}[t]
	\centering
	\includegraphics[width=\textwidth]{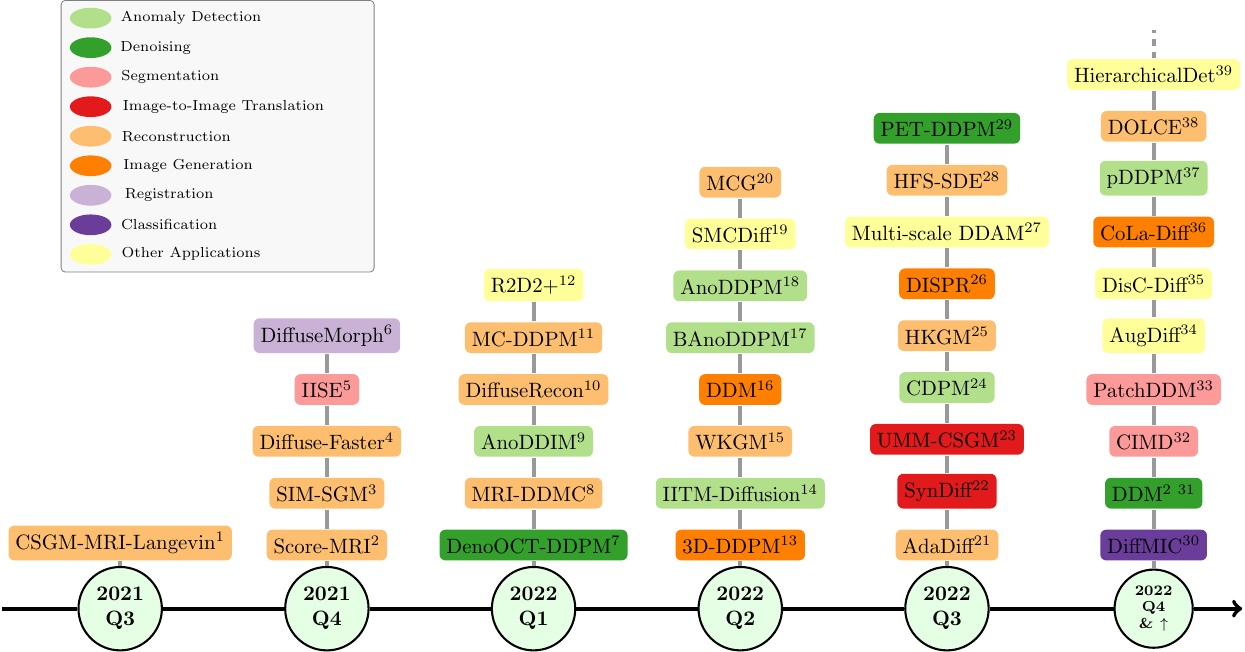}
	\caption{Diffusion models timeline from 2021 to April 2023 through the first paper in the medical field. The superscripts in ascending order represent 1.\cite{jalal2021robust}, 2.\cite{chung2022score}, 3.\cite{song2022solving}, 4.\cite{chung2022come}, 5.\cite{wolleb2022diffusion}, 6.\cite{kim2022diffusemorph}, 7.\cite{hu2022unsupervised}, 8.\cite{luo2022mri}, 9.\cite{wolleb2022diffusionanomaly}, 10.\cite{peng2022towards}, 11.\cite{xie2022measurement}, 12.\cite{chung2022mr}, 13.\cite{dorjsembe2022three}, 14.\cite{wolleb2022swiss}, 15.\cite{tu2022wkgm}, 16.\cite{kim2022diffusion}, 17.\cite{pinaya2022fast}, 18.\cite{wyatt2022anoddpm}, 19.\cite{trippe2023diffusion}, 20.\cite{chung2022improving}, 21.\cite{dar2022adaptive}, 22.\cite{ozbey2022unsupervised}, 23.\cite{meng2022novel}, 24.\cite{sanchez2022healthy}, 25.\cite{peng2022one}, 26.\cite{waibel2022diffusion}, 27.\cite{wang2022fight}, 28.\cite{cao2022high}, 29.\cite{gong2022pet}, 30.\cite{yang2023diffmic}, 31.\cite{xiang2023ddm}, 32.\cite{rahman2023ambiguous}, 33.\cite{bieder2023diffusion}, 34.\cite{shao2023augdiff}, 35.\cite{mao2023disc}, 36.\cite{jiang2023cola}, 37.\cite{behrendt2023patched}, 38.\cite{liu2022dolce}, 39.\cite{hamamci2023diffusion}, respectively.}
	\label{fig:timeline}
\end{figure*}

As seen in \Cref{fig:timeline} and \Cref{fig:taxonomy}, some specific applications, such as image reconstruction, denoising, and generation, have received more attention compared to other tasks like segmentation, text-to-image translation, registration, etc. This is largely due to the compatibility of the diffusion model theory with the objectives of reconstruction and denoising tasks. The process in diffusion models involves adding noise to data and then denoising it until the original data is reconstructed, making it easier to implement these two tasks within the framework of diffusion models. Moreover, diffusion models have the ability to capture the underlying physical processes involved in these tasks and effectively model the complex interactions between signals and noise in data, resulting in more accurate image reconstruction. In addition, diffusion models are a class of generative models that operate on probabilistic distributions and can be conditioned to create synthetic data with a high degree of diversity and quality, which is why image generation has also been a popular application from the start.

While diffusion models have the potential to be applied to different tasks, they may require further modifications to be adapted to other specific tasks. For example, text-to-image translation requires an auxiliary network with strong text encoding capabilities. In the case of object detection, recent vision-based work has demonstrated the potential of applying diffusion models to the object detection task by progressively refining randomly generated boxes to produce the final output results \cite{chen2022diffusiondet}. Hence, while initial works tend to focus on image generation, reconstruction, and denoising tasks, it is expected that, over time, more research will emerge addressing a wider range of tasks, as an examination of \Cref{fig:timeline} reveals the promising future of this field in the academic environment.
\begin{table*}[!h]
    \centering
    \caption{Overview of the reviewed diffusion models in medical imaging based on their algorithm choice presented in our taxonomy, \Cref{fig:taxonomy}. The symbol * indicates that the mentioned paper explores both DDPM and SDE algorithms.}
    \label{tab:paperhighlights}
    \resizebox{\textwidth}{!}{
    \begin{tabular}{lp{6cm}p{8cm}p{7cm}} 
    \hline
    \rowcolor[rgb]{0.945,0.804,0.922}   & & &  \\ 
    \rowcolor[rgb]{0.945,0.804,0.922}   \vcell{\textbf{Algorithm}} & \vcell{\textbf{Networks}} & \vcell{\textbf{Core Ideas}} & \vcell{\textbf{Practical Use Cases}} \\[-\rowheight]
    \rowcolor[rgb]{0.945,0.804,0.922}   \printcelltop & \printcelltop & \printcelltop & \printcelltop \\ 
    \toprule
    \vcell{Denoising Diffusion Probabilistic Models (DDPMs)} & 
    \vcell{
    $^{1}$AnoDDPM \cite{wyatt2022anoddpm}\newline
    $^{2}$CDPM \cite{sanchez2022healthy}\newline
    $^{3}$IITM-Diffusion \cite{wolleb2022swiss}\newline
    $^{4}$AnoDDIM \cite{wolleb2022diffusionanomaly}\newline
    $^{5}$PET-DDPM \cite{gong2022pet}\newline
    $^{6}$DenoOCT-DDPM \cite{hu2022unsupervised}\newline
    $^{7}$brainSPADE \cite{fernandez2022can}\newline
    $^{8}$DARL \cite{kim2022diffusion}\newline
    $^{9}$IISE \cite{wolleb2022diffusion}\newline 
    $^{10}$*MT-Diffusion \cite{lyu2022conversion}\newline
    $^{11}$SynDiff \cite{ozbey2022unsupervised}\newline
    $^{12}$MC-DDPM \cite{xie2022measurement}\newline
    $^{13}$DiffuseRecon \cite{peng2022towards}\newline
    $^{14}$AdaDiff \cite{pinaya2022brain}\newline
    $^{15}$BrainGen \cite{dar2022adaptive}\newline
    $^{16}$MFDPM \cite{moghadam2023morphology}\newline
    $^{17}$DISPR \cite{waibel2022diffusion}\newline
    $^{18}$3D-DDPM \cite{dorjsembe2022three}\newline
    $^{19}$DDM \cite{kim2022diffusion}\newline
    $^{20}$Multi-scale DDAM \cite{wang2022fight}\newline
    $^{21}$SMCDiff \cite{trippe2023diffusion}\newline
    $^{22}$BAnoDDPM \cite{pinaya2022fast} \newline
    $^{23}$DiffuseMorph \cite{kim2022diffusemorph} \newline
    $^{24}$20x-DenoDDPM \cite{xia2022low} \newline
    $^{25}$DOLCE \cite{liu2022dolce} \newline
    $^{26}$DiffMIC \cite{yang2023diffmic} \newline
    $^{27}$CIMD \cite{rahman2023ambiguous} \newline
    $^{28}$PatchDDM \cite{bieder2023diffusion} \newline
    $^{29}$DDM$^2$ \cite{xiang2023ddm} \newline
    $^{30}$CoLa-Diff \cite{jiang2023cola} \newline
    $^{31}$pDDPMs \cite{behrendt2023patched} \newline
    $^{32}$MLS-DDPM \cite{gong2023diffusion} \newline
    $^{33}$X-ray LDM \cite{packhauser2022generation} \newline
    $^{34}$DAE \cite{keicher2023semantic} \newline
    $^{35}$HierarchicalDet \cite{hamamci2023diffusion}} 
     & 
    \vcell{In DDPMs \cite{ho2020denoising}, the forward diffusion process is represented as a Markov chain in which Gaussian noise is gradually added to the data. Data generation is then accomplished using the attained pure random noise and begins iterative denoising through a parametrized reverse process. Unlike VAE, where both the encoder and decoder are trained, only a single network is trained during the reverse process, and the forward process is considered fixed. An objective function of DDPMs is to simulate noise, which means that given a noisy input image, the neural network will produce the distribution modeled as normal distribution and indicate where the noise originates.} & 
    \vcell{
    $\bullet$ Generalization to other types of noise distributions \cite{wyatt2022anoddpm} \newline  
    $\bullet$ Facilitating diffusion process via implicit guidance \cite{sanchez2022healthy} \newline
    $\bullet$ Acceleration improvement using DDIM sampling \cite{wolleb2022swiss,keicher2023semantic,wolleb2022diffusionanomaly} \newline
    $\bullet$ Guiding diffusion process via classifier guidance \cite{wolleb2022diffusionanomaly} \newline
    $\bullet$ Conditional DDPMs \cite{gong2022pet,lyu2022conversion,xie2022measurement,waibel2022diffusion,xia2022low,trippe2023diffusion} \newline
    $\bullet$ Generating synthetic segmentation datasets \cite{fernandez2022can,kim2022vessel} \newline
    $\bullet$ Cross-modality translation \cite{ozbey2022unsupervised} \newline
    $\bullet$ Multi-modal conversion \cite{lyu2022conversion, ozbey2022unsupervised} \newline
    $\bullet$ Exploiting K-space parameter-free guidance \cite{peng2022towards} \newline
    $\bullet$ Accelerate MC sampling using coarse-to-fine sampling \cite{peng2022towards} \newline
    $\bullet$ Adversarial learning in the reverse diffusion process \cite{dar2022adaptive,ozbey2022unsupervised} \newline
    $\bullet$ 3D reconstruction from 2D images \cite{waibel2022diffusion} \newline
    $\bullet$ Conditioning on medical meta-data \cite{pinaya2022brain} \newline
    $\bullet$ Using LDM to enhance the training and sampling efficiency \cite{pinaya2022brain,jiang2023cola,pinaya2022fast,packhauser2022generation}\newline
    $\bullet$ DDPMs for histopathology images \cite{moghadam2023morphology}\newline
    $\bullet$ 3D medical image generation \cite{dorjsembe2022three} \newline
    $\bullet$ Using deformation fields for medical image generation \cite{kim2022diffusemorph,kim2022diffusion} \newline
    $\bullet$ DDPMs in skin image adversarial attacks \cite{wang2022fight} \newline
    
    $\bullet$ CT image reconstruction using limited-angle sinograms \cite{liu2022dolce} \newline
    $\bullet$ Improving efficiency using a patch-based strategy \cite{bieder2023diffusion,behrendt2023patched} \newline
    $\bullet$ Denoising diffusion MRI \cite{xiang2023ddm}}
     \\[-\rowheight]
    \printcelltop & \printcelltop & \printcelltop & \printcelltop \\ 
    \midrule
    \vcell{Noise Conditioned Score Networks (NCSNs)} 
    & 
    \vcell{
    $^{1}$CSGM-MRI-Langevin \cite{jalal2021robust} \newline
    $^{2}$Self-Score \cite{cui2022self}
    } 
    & 
    \vcell{
    In this algorithm \cite{song2019generative}, creating samples requires solving the Langevin dynamics equation. However, this equation mandates the solution of the gradient of the log density w.r.t. the input, $\nabla_x \log p(x)$, which is unknown and intractable. NCSN formulates the forward diffusion process by disturbing the data with Gaussian noise at different scales. Through this approach, this equation can be solved by training a single score network conditioned on the noise level and modeling the scores at all noise levels. Therefore, using the annealed Langevin dynamics algorithm and estimated score function at each scale, data can be generated.
    } 
    & 
    \vcell{
    $\bullet$ Using Langevin dynamics with different random initializations to get multiple reconstructions \cite{jalal2021robust} \newline
    $\bullet$ Using conditioned Langevin Markov chain Monte Carlo (MCMC) sampling \cite{cui2022self} \newline
    $\bullet$ Utilizing K-space data \cite{cui2022self}
    } \\[-\rowheight]
    \printcelltop & \printcelltop & \printcelltop & \printcelltop \\ 
    \midrule
    \vcell{Stochastic Differential Equations (SDEs)} 
    & 
    \vcell{$^{1}$*MT-Diffusion \cite{lyu2022conversion}\newline
     $^{2}$UMM-CSGM \cite{meng2022novel}\newline
     $^{3}$SIM-SGM \cite{song2022solving}\newline
     $^{4}$Score-MRI \cite{chung2022score}\newline
     $^{5}$MRI-DDMC \cite{luo2022mri}\newline
     $^{6}$MCG \cite{chung2022improving}\newline
     $^{7}$HFS-SDE \cite{cao2022high}\newline
     $^{8}$Diffuse-Faster \cite{chung2022come}\newline
     $^{9}$HKGM \cite{peng2022one} \newline
     $^{10}$WKGM \cite{tu2022wkgm}\newline
     $^{11}$R2D2+ \cite{chung2022mr}
     }
    & 
    \vcell{As in the cases of DDPMs and NCSNs, SDEs \cite{song2020score} follow a similar approach in the forward diffusion process, in which the input data is perturbed consecutively utilizing Gaussian noise. In contrast, previous probabilistic models can be viewed as discretizations of score-based models by extending the number of noise scales to infinity and treating them as continuous. \newline
    Having operated backward in time, the reverted SDE Process solves the reverse-time Stochastic Differential Equation and recovers the data. Even so, each time step of this process requires the score function of the density. SDEs are, therefore, intended to learn the actual score function via a neural network and construct samples using numerical SDE solvers. \newline
    Any numerical method can be adopted to solve the reverse-time SDE equation. In particular, three notable sampling methods are named and described in \Cref{SDE-sampling-methods}: Euler-Maruyama (EM) method, Prediction-Correction (PC), and Probability Flow ODE. EM follows a simple strategy, PC generates more high-fidelity samples, and Probability Flow ODE is fast and efficient.} 
    & 
    \vcell{ 
    $\bullet$ SDEs achieve better results in multi-modal conversion \cite{lyu2022conversion} \newline
    $\bullet$ Cross-modality translation \cite{meng2022novel} \newline
    $\bullet$ Conditional SDEs \cite{meng2022novel,peng2022towards,chung2022improving} \newline
    $\bullet$ Solving linear inverse problems \cite{song2022solving} \newline
    $\bullet$ Using K-space data \cite{peng2022one,tu2022wkgm} \newline
    $\bullet$ Using manifold constraints \cite{chung2022improving} \newline
    $\bullet$ Good initialization in the reverse process instead of random Gaussian noise leads to a faster convergence \cite{chung2022come,cao2022high} \newline
    $\bullet$ Diffusion only in high-frequency space increases the stability and quality \cite{cao2022high} \newline
     }\\[-\rowheight]
    \printcelltop & \printcelltop & \printcelltop & \printcelltop \\
    \bottomrule
    \end{tabular}
    }
\end{table*}

\section{Future Direction and Open Challenges}
\label{future-direction}
Diffusion models have emerged as a popular topic in the medical vision and medical-biology fields, as evidenced by the upward trend shown in \Cref{fig:charts}. One of the primary advantages of diffusion models in medical imaging is that they do not require labeled data, making them a strong candidate for many medical applications. In addition, Diffusion models have gained popularity due to their impressive performance in foundational models like large text-to-image models \cite{ramesh2022hierarchical}. These models remain appealing for modeling images and other data distributions for several reasons, including their ability to effectively represent high-dimensional data like images due to their inductive bias. These models are trained on a reweighted variational lower bound that emphasizes the global consistency and dominant patterns of images while giving less attention to less noticeable details, making them an excellent inductive bias for spatial data. However, Diffusion models, compared to other generative models, have some limitations that should be taken into account when considering their applications. These limitations include a slower generation process compared to some other generative models, limited applicability to certain data types (e.g., audio, text, or structured data), lower likelihood compared to other models, and an inability to perform dimensionality reduction \cite{cao2022survey}. However, these limitations do not diminish the unique strengths and advantages of diffusion models in generating high-quality images and their ability to work without a pair of labeled or unlabeled data but rather present open challenges for further research and improvements.

This paper aims to provide a comprehensive review of the latest medical research papers utilizing diffusion models. We categorized the studies based on the taxonomy proposed in \Cref{fig:taxonomy} to demonstrate the potential of diffusion models. Through this review, we hope to highlight the power of diffusion models and shed light on their importance in advancing the capabilities of medical imaging techniques. This section identifies areas for future investigation, emphasizing the need for continued research in this exciting and rapidly evolving field.

\noindent\textbf{Exploring more diverse medical imaging modalities:}
Due to the nature of diffusion models, they are a strong candidate for exploring diverse modalities for distinct downstream tasks. According to \Cref{fig:modalities}, most of the published studies utilize CT and MRI as modalities. Nevertheless, other modalities may also benefit from the capabilities of diffusion models. For instance, ultrasound imaging techniques may suffer from a non-ideal Point Spread Function (PSF) of the imaging system as well as intrinsic physical limitations \cite{goudarzi2022deep}. To this end, several studies explored the impact of various generative pipelines on ultrasound data in image quality enhancement and denoising \cite{goudarzi2022deep,zhang2022ultrasound,van2021ultrasound}, resulting in improved image quality.

\noindent\textbf{Representation space:} 
VAEs and GANs are designed to preserve and learn meaningful representations of data in their latent space. Nevertheless, diffusion models have been proven to be less successful in creating semantically 
meaningful data representations in their latent space \cite{yang2022diffusion,traub2022representation}. Hence, the lack of semantically expressive data representations in the latent space of diffusion models poses a significant hindrance in performing tasks that involve the manipulation of data based on semantic representations. This is probably because diffusion models mainly destruct information in the latent variables during the diffusion process, resulting in less meaningful representation space. Therefore, there is a need to develop models that can learn semantically meaningful representations, as they enable better image reconstructions and semantic interpolations \cite{traub2022representation}. For example, Abstreiter et al. \cite{abstreiter2021diffusion} have proposed a novel diffusion-based representation learning approach based on conditional denoising score matching. Specifically, they introduce an additional trainable encoder and condition the score estimator on the encoder output, which is the latent representation of the clean data. This produces interpretable features in the latent space and enables altering the encoded features' granularity without requiring architecture modifications or data manipulations. Therefore, the lack of proper representation space of diffusion models presents an open challenge for researchers to work on.


\noindent\textbf{Architecture design:} 
In the context of diffusion models, the network structure is a critical design choice that directly affects their ability to learn complex data relationships and produce high-quality results on large and diverse datasets \cite{dhariwal2021diffusion}. Most diffusion models currently utilize CNN-based architectures with a global attention layer, but recent studies have explored the use of transformer models \cite{peebles2022scalable}. Compared to CNNs, transformers offer several advantages, including the ability to model non-local interactions and capture long-range dependencies within the data. These models have also shown promising results in natural language processing tasks, indicating their potential for modeling sequential or spatiotemporal data. Despite this potential, the use of transformers in diffusion models is still in its early stages, and further research is necessary to fully understand their capabilities and limitations in this context. Specifically, most DDPM-based approaches follow the baseline of \cite{ho2020denoising,dhariwal2021diffusion,nichol2021improved}, while Score-based approaches follow the baseline of \cite{song2020score,song2020improved}. As a result, there is a lack of research focused on improving the architecture of diffusion models for medical imaging, making it an open challenge for future research.

\noindent\textbf{Causal discovery, inference, and counterfactual generation of diffusion models:} 
Diffusion models can generally learn the underlying probability distribution of a dataset and be used to generate new data points that follow the same distribution. This makes them even more useful for complex causal inference, discovery, and counterfactual generation tasks. The benefit of using diffusion models for causal inference is their ability to handle missing data and their robustness to distributional shifts, allowing them to estimate the causal effects of interventions in real-world settings where data may be incomplete. For example, Sanchez et al. \cite{sanchez2022diffusion} propose a novel framework for estimating causal effects involving high-dimensional variables using diffusion models. In addition, causal discovery seeks to identify the underlying causal structure of a system without necessarily being given a specific intervention. Sanchez et al. \cite{sanchez2023diffusion} proposed a novel method using diffusion models for causal discovery via topological ordering based on diffusion models. Additionally, diffusion models can be used to generate counterfactuals, which explore hypothetical scenarios that assess the impact of an intervention that was or was not made. This is particularly useful in fields such as medicine and public policy, where conducting randomized controlled trials may be difficult or unethical. Overall, diffusion models provide a powerful tool for generating data and conducting causal inference, discovery, and counterfactual analysis in various fields.

\noindent\textbf{Privacy concerns:}
The medical community is highly concerned about the privacy of medical data. AI image synthesis models are currently under scrutiny for potentially violating copyright laws and compromising the privacy of their training data. Carlini et al. \cite{carlini2023extracting} conducted extensive experiments to evaluate the privacy concerns associated with generative diffusion models. The results show that diffusion models tend to memorize individual images from their training data and reproduce them during generation. This would consequently enable adversaries to attack to extract training data. Furthermore, the study reveals that diffusion models are much less private compared to other generative models like GANs. As a result, new developments in privacy-preserving training are needed to handle these vulnerabilities, particularly in delicate fields like medicine.

\noindent\textbf{Federated learning and diffusion models:}
Due to privacy concerns in medical imaging, which limit data integration, diffusion models and federated learning can create a profound and robust learning platform in the medical domain. Data is collected from various sources in privacy preservation smart healthcare systems and stored in decentralized locations \cite{ali2022federated}. Federated Learning, as it allows for training machine learning models on decentralized data without exposing sensitive information in conjunction with diffusion models, can capture the underlying distribution of the data across multiple participating devices. This intuition, using the diffusion models as a ``generative prior," can help to mitigate the effects of data heterogeneity, reduce the risk of privacy leaks, and improve the quality of the learned models and their trustworthiness in fairness generalization. Furthermore, generative models (i.e., diffusion models) are trained to learn the underlying probability distribution of the data rather than memorizing the training data. The stability against perturbations is a desirable property of machine learning models, especially in safety-critical applications, such as medical diagnosis, where small changes in the input data can have significant consequences. Diffusion models can provide stability against perturbations by using regularization techniques, such as adding noise to the input data during training, which can help the model generalize better to unseen data \cite{jalal2021robust,chung2022score}. Excessive to this, federated learning models perform on a broadened range of sources with various data distributions. Therefore, a federated learning paradigm could achieve strong out-of-distribution (OOD) generalization capacity \cite{cui2022self,chung2022score}. In addition, due to the same privacy concerns, diffusion models can consolidate their steps in generating synthetic medical data for educational purposes.

\noindent\textbf{Reverse process using reinforcement learning:} The inverse problem-solving of the diffusion models could be performed by the reinforcement learning paradigm to estimate the best inversion path rather than solid mathematical solutions. In this process, reinforcement learning can be used to search for the optimal values of the diffusion model parameters that maximize a given reward function. The reward function can be designed to evaluate how well the diffusion model fits the data and penalize deviations from the observed data where the traditional optimization methods, such as maximum likelihood estimation or Bayesian inference, become computationally expensive or intractable.

\section{Conclusion}
\label{conclusion}
In this paper, we provided a survey of the literature on diffusion models with a focus on applications in the medical imaging field. Specifically, we investigated the applications of diffusion models in anomaly detection, medical image segmentation, denoising, classification, reconstruction, registration, generation, and other tasks. In particular, for each of these applications, we provided a taxonomy and high-level abstraction of the core techniques from various angles. Moreover, we characterized the existing models based on techniques where we identified three primary formulations of diffusion modeling based on: DDPMs, NCSNs, and SDEs. Finally, we outlined possible avenues for future research.

While our survey highlights the rapid growth of diffusion-based techniques in medical imaging, we also acknowledge that the field is still in its early stages and subject to change. As diffusion models continue to gain popularity and more research is conducted in this field, our survey serves as an important starting point and reference for researchers and practitioners looking to utilize these models in their work. We hope that this survey will inspire further interest and exploration of the potential of diffusion models in the medical domain. It is important to note that some of the papers cited in this survey are pre-prints. However, we made every effort to include only high-quality research from reputable sources, and we believe that the inclusion of pre-prints provides a comprehensive overview of the current state-of-the-art in this rapidly evolving field. Overall, we believe that our survey provides valuable insights into the use of diffusion models in medical imaging and highlights promising areas for future research.

\bibliographystyle{unsrt}
\bibliography{Ref}

\end{document}